\documentclass{article}%
\usepackage{graphicx}
\usepackage{amsmath}
\usepackage{amsfonts}
\usepackage{amssymb}%
\newtheorem{theorem}{Theorem}
{}

\newtheorem{lemma}{Lemma}
{}

\newtheorem{remark}{Remark}

\newenvironment{proof}[1][Proof]{\textbf{#1.} }{\ \rule{0.5em}{0.5em}}

\begin{document}

\author{O. A. Veliev\\{\small \ Depart. of Math., Faculty of Arts and Sci., Dogus University,}\\{\small Acibadem, Kadikoy, Istanbul, Turkey,}\\{\small \ e-mail: oveliev@dogus.edu.tr}}
\title{On the Constructively Determination the Spectral Invariants of the Periodic
Multidimensional Schr\"{o}dinger Operator}
\date{}
\maketitle

\begin{abstract}
In this paper we constructively determine a family of the spectral invariants
of the multidimensional Schr\"{o}dinger operator with a periodic potential by
the given band functions.

\end{abstract}

\section{Introduction}

We investigate the Schr\"{o}dinger operator
\begin{equation}
L(q)=-\Delta+q(x),\text{ }x\in\mathbb{R}^{d},\text{ }d\geq2
\end{equation}
with a real periodic (relative to the lattice $\Omega$) potential $q(x)\in
W_{2}^{s}(F),$ where $s\geq6(3^{d}(d+1)^{2})+d$ and $F$ is the fundamental
domain $\mathbb{R}^{d}/\Omega$ of $\Omega.$ The spectrum of $L(q)$ is the
union of the spectra of the operators $L_{t}(q)$ for $t\in F^{\ast}%
\equiv\mathbb{R}^{d}/\Gamma$ generated by (1) and the conditions
\[
u(x+\omega)=e^{i(t,\omega)}u(x),\ \forall\omega\in\Omega,
\]
where $\Gamma\equiv\{\delta\in\mathbb{R}^{d}:(\delta,\omega)\in2\pi
\mathbb{Z},\forall\omega\in\Omega\}$ is the lattice dual to $\Omega$ ( see
[1]). The eigenvalues $\Lambda_{1}(t)\leq\Lambda_{2}(t)\leq...$of $L_{t}(q)$
define functions $\Lambda_{1}(t),$ $\Lambda_{2}(t),...,$ of $t$ that are
called the band functions of $L(q)$. In this paper using the asymptotic
formulas for the band functions and the Bloch functions obtained in [4], we
obtain \ more detailed asymptotic formulas and then constructively determine a
family of the spectral invariants by the given band functions. In introduction
we list the main results. In section 2 we prove the main results without
giving some estimations which are given in section 3 and in appendices.

Let $\delta$ be a maximal element of $\Gamma$, that is, $\delta$ is the
nonzero element of $\Gamma$ of minimal norm belonging to the line
$\delta\mathbb{R}$ and
\begin{equation}
q^{\delta}(x)=\sum_{n\in Z}q_{n\delta}e^{in(\delta,x)}=Q(\zeta)
\end{equation}
be the directional (one dimensional) potential, where $\zeta=(\delta,x)$ and%
\[
q_{\gamma}=(q(x),e^{i(\gamma,x)})=\int_{F}q(x)e^{-i(\gamma,x)}dx
\]
is the Fourier coefficient of $q(x).$ Without loss of generality we assume
that the measure $\mu(F)$ of $F$ is $1$ and $q_{0}=0.$ Let $\lambda_{0}%
\leq\lambda_{2}^{-}\leq\lambda_{2}^{+}...$ and

$\lambda_{1}^{-}\leq\lambda_{1}^{+}\leq\lambda_{3}^{-}\leq\lambda_{3}^{+}%
...$be the eigenvalues of the boundary value problem%

\begin{equation}
-\mid\delta\mid^{2}y^{\prime\prime}(\zeta)+Q(\zeta)y(\zeta)=\mu y(\zeta
),\text{ }y(\zeta+2\pi)=e^{i2\pi v}y(\zeta)
\end{equation}
for $v=0$ and $v=\frac{1}{2}$ respectively, where $\mid\delta\mid$ is the norm
of $\delta.$ The corresponding eigenfunctions are denoted by $\varphi_{0}(s)$
and $\varphi_{n}^{\pm}(s)$ respectively.

In the pioneering paper [2] about isospectral potentials it was proved that if
$\ q(x)\in C^{6}(F),$ $\omega\in\Omega\backslash0,$ and $\delta$ is the
maximal element of $\Gamma$ satisfying $(\delta,\omega)=0$ then given band
functions one may recover $\lambda_{0},\lambda_{1}^{-},\lambda_{1}^{+}%
,\lambda_{2}^{-},\lambda_{2}^{+},...$ and%
\[
\int_{F}\mid Q_{\omega}(x)\varphi_{n}^{\pm}(s)\mid^{2}dx\text{ \ if \ }%
\lambda_{n}^{-}<\lambda_{n}^{+},
\]

or $\ \int_{F}\mid Q_{\omega}(x)\mid^{2}((\varphi_{n}^{+}(s))^{2}+(\varphi
_{n}^{-}(s))^{2})dx$ \ if $\ \lambda_{n}^{-}=\lambda_{n}^{+},$ where
\[
Q_{\omega}(x)=\sum_{\gamma:\gamma\in\Gamma,(\gamma,\omega)\neq0}\frac{\gamma
}{(\omega,\gamma)}q_{\gamma}e^{i(\gamma,x)}.
\]
The proofs given there were nonconstructive. In paper [3] it was given a
constructive way of determining the spectrum of $L_{t}(q^{\delta})$\ from the
spectrum of $L_{t}(q)$ for the two dimensional ($d=2$ ) case (see remark (1)
of [3]).

In this paper, for arbitrary dimension $d,$ by the given band functions we
constructively determine the all eigenvalues of the boundary value problem (3)
for all values of $v$ and a family of new spectral invariants
\[
J(\delta,b,n,v),\text{ }J_{0}(\delta,b),\text{ }J_{1}(\delta,b),\text{ }%
J_{2}(\delta,b)\text{ \ \ \ ( see (12), (15))}%
\]
for $\upsilon\in(0,\frac{1}{2})\cup(\frac{1}{2},1)$, $n\in\mathbb{Z}$,
$\delta\in M(\Gamma),$ $b\in M(\Gamma_{\delta}),$ where $M(\Gamma)$ and
$M(\Gamma_{\delta})$ are the set of all maximal elements of the lattices
$\Gamma$ and $\Gamma_{\delta}$ respectively, $\Gamma_{\delta}$ is the dual
lattice of $\Omega_{\delta}$ and $\Omega_{\delta}=\{h\in\Omega:(h,\delta)=0\}$
is the sublattice of $\Omega$ in the hyperplane $H_{\delta}=$ $\{x\in
\mathbb{R}^{d}:(x,\delta)=0\}$. Note that $J_{k}(\delta,b)$ is explicitly
expressed by Fourier coefficient of $q(x).$ Moreover, if $d>2$ and $q(x)$ is a
trigonometric polynomial then, in general, the number of nonzero spectral
invariants $J_{k}(\delta,b)$ is greater than the number of nonzero Fourier
coefficient of $q(x).$This situation allows us to give ( it will be given in
next papers) an algorithm for finding the potential $q(x)$ from these spectral invariants.

Let us describe the brief scheme of this paper. First using the asymptotic
formulas for the band functions and the Bloch functions obtained in [4], we
obtain \ more detailed asymptotic formulas and then using these formulas we
constructively determine the family of the spectral invariants. The
eigenvalues of the operator $L_{t}(0)$ with zero potential are $\mid
\gamma+t\mid^{2}$ for $\gamma\in\Gamma.$ If the quasimomentum $\gamma
+t\ $lies\ near the diffraction plane
\begin{equation}
D_{\delta}=\{x\in\mathbb{R}^{d}:\mid x\mid^{2}-\mid x+\delta\mid^{2}=0\},
\end{equation}
then the corresponding eigenvalue of $L_{t}(q)$ is close to the eigenvalue of
the operator $L_{t}(q^{\delta})$ with directional potential (2). To describe
the eigenvalue of $L_{t}(q^{\delta})$ we consider the lattice $\Gamma_{\delta
}.$ Let $F_{\delta}\equiv H_{\delta}/\Gamma_{\delta\text{ }}$ be the
fundamental domain of $\Gamma_{\delta}.$ In this notation the quasimomentum
$\gamma+t$ has the orthogonal decompositions
\begin{equation}
\gamma+t=\beta+\tau+(j+v)\delta,
\end{equation}
where $\beta\in\Gamma_{\delta}\subset H_{\delta},$ $\tau\in F_{\delta}\subset
H_{\delta},$ $j\in\mathbb{Z}$, $v\in\lbrack0,1)$ and $v$ depends on $\beta$
and $t.$ The eigenvalues and eigenfunctions of the operator $L_{t}(q^{\delta
})$ are
\begin{equation}
\lambda_{j,\beta}(v,\tau)=\mid\beta+\tau\mid^{2}+\mu_{j}(v),\text{ }%
\Phi_{j,\beta}(x)=e^{i(\beta+\tau,x)}\varphi_{j,v}(\zeta)
\end{equation}
for $j\in\mathbb{Z},$ $\beta\in\Gamma_{\delta},$ where $\mu_{j}(v)$ and
$\varphi_{j,v}(\zeta)$ are eigenvalues and eigenfunctions of the operator
$T_{v}(Q)$ generated by the boundary value problem (3). We say that the large
quasimomentum (5) lies near the diffraction plane (4) if
\begin{equation}
\frac{1}{2}\rho<\mid\beta\mid<\frac{3}{2}\rho,\text{ }j=O(\rho^{\alpha_{1}}),
\end{equation}
where $\rho$ is large parameter, $\alpha=\frac{1}{4(3^{d}(d+1))},$ and
$\alpha_{k}=3^{k}\alpha$ for $k=1,2,...,d.$ In this paper we construct a set
of quasimumentum near the diffraction plane $D_{\delta}$ such that if
$\ \beta+\tau+(j+v)\delta$ ( see (5)) belongs to this set, then there exists a
simple eigenvalue, denoted by $\Lambda_{j,\beta}(v,\tau),$ of $L_{t}(q)$
satisfying
\begin{equation}
\Lambda_{j,\beta}(v,\tau)=\lambda_{j,\beta}(v,\tau)+O(\rho^{-a}),
\end{equation}%
\begin{equation}
\Lambda_{j,\beta}(v,\tau)=\lambda_{j,\beta}(v,\tau)+\frac{1}{4}\int_{F}\mid
f_{\delta,\beta+\tau}^{2}\mid\left\vert \varphi_{j,v}\right\vert ^{2}%
dx+O(\rho^{-3a+2\alpha_{1}}\ln\rho),
\end{equation}
where $a=1-\alpha_{d}+\alpha$ and%
\[
f_{\delta,\beta+\tau}(x)=\sum_{\gamma:\gamma\in\Gamma\backslash\delta
\mathbb{R},\mid\gamma\mid<\rho^{\alpha}\text{ }}\frac{\gamma}{(\beta
+\tau,\gamma)}q_{\gamma}e^{i(\gamma,x)}.
\]
The eigenfunction $\Psi_{j,\beta}(x)$ corresponding to $\Lambda_{j,\beta
}(v,\tau)$ satisfies
\begin{equation}
\Psi_{j,\beta}(x)=\Phi_{j,\beta}(x)+O(\rho^{-a}).
\end{equation}
Besides we prove that derivative of $\Lambda_{j,\beta}(v,\tau)$ in direction
$h=\frac{\beta+\tau}{\mid\beta+\tau\mid}$ satisfies
\begin{equation}
\mid\beta+\tau\mid\frac{\partial\Lambda_{j,\beta}(v,\tau)}{\partial h}%
=\mid\beta+\tau\mid^{2}+O(\rho^{2-2a})
\end{equation}
and the derivative of other simple eigenvalues, neighboring with
$\Lambda_{j,\beta}(v,\tau),$ does not satisfy (11). Using this formulas we
constructively determine the eigenvalues $\mu_{n}(v)$ for $n\in\mathbb{Z}$,
$v\in\lbrack0,1)$ and the spectral invariants
\begin{equation}
\text{ }J(\delta,b,n,v)=\int_{F}\mid q_{\delta,b}(x)\varphi_{n,v}%
(\delta,x)\mid^{2}dx
\end{equation}
for $\upsilon\in(0,\frac{1}{2})\cup(\frac{1}{2},1)$, $n\in\mathbb{Z}$, and for
all maximal elements $b$ of $\Gamma_{\delta},$ where $\delta$ is any maximal
element of $\Gamma,$
\begin{equation}
q_{\delta,b}(x)=\sum_{\gamma\in S(\delta,b)\backslash\delta\mathbb{R}}%
\frac{\gamma}{(b,\gamma)}q_{\gamma}e^{i(\gamma,x)},
\end{equation}
$S(\delta,b)=P(\delta,b)\cap\Gamma,$ and $P(\delta,b)$ is the plane containing
$\delta$, $b$ and $0$. Then substituting the asymptotic decomposition
\begin{equation}
\left\vert \varphi_{n,v}(\zeta)\right\vert ^{2}=A_{0}+\frac{A_{1}(\zeta)}%
{n}+\frac{A_{2}(\zeta)}{n^{2}}+...,
\end{equation}
where $A_{k}(\zeta)$ is expressed via $Q(\zeta)$ ( see (2)), into (12) we find
the invariants
\begin{equation}
J_{k}(\delta,b)=\int_{F}|q_{\delta,b}(x)|^{2}A_{k}(\zeta)dx
\end{equation}
for $k=0,1,2,...$.Using the well-known asymptotic formulas for eigenvalues and
eigenfunctions of the Sturm-Liouville operator $T_{v}(Q)$ by direct
calculations we find $A_{0}(\zeta),$ $A_{1}(\zeta),$ $A_{2}(\zeta)$ and the
invariants
\begin{equation}
\int_{F}\left\vert q_{\delta,b}(x)\right\vert ^{2}q^{\delta}(x)dx,
\end{equation}%
\begin{equation}
\int_{F}\left\vert q^{\delta}(x)\right\vert ^{2}dx
\end{equation}
( see Appendix D). If the potential $q(x)$ is a trigonometric polynomial then
the most of the directional potentials has the form
\begin{equation}
q^{\delta}(x)=q_{\delta}e^{i(\delta,x)}+q_{-\delta}e^{-i(\delta,x)}.
\end{equation}
In this case, by direct calculations, we show that (see Appendix D)
\begin{align}
A_{0}  &  =1,\text{ }A_{1}=0,\text{ }A_{2}=\frac{q^{\delta}(x)}{2}%
+a_{1}\left\vert q_{\delta}\right\vert ^{2},A_{3}=a_{2}q^{\delta}%
(x)+a_{3}\left\vert q_{\delta}\right\vert ^{2},\\
A_{4}  &  =a_{4}q^{\delta}(x)+a_{5}(q_{\delta}^{2}e^{i2(\delta,x)}+q_{-\delta
}^{2}e^{-i2(\delta,x)})+a_{6},\nonumber
\end{align}
where $a_{1},a_{2},...,a_{6}$ are the known constants. Moreover using (19),
(17), and (15) for $k=2,4$ we find the invariant
\begin{equation}
\int|q_{\delta,b}(x)|^{2}(q_{\delta}^{2}e^{i2(\delta,x)}+q_{-\delta}%
^{2}e^{-i2(\delta,x)})dx
\end{equation}
in the case (18). In next paper we give an algorithm for finding the potential
$q(x)$ by the invariants (16), (17), and (20).

\section{The Proofs of the Main Results}

In this section we give the proofs of the main results without getting the
technical details. The technical details, namely the proof of lemmas and some
estimations are investigated in Sections 3 and in appendices respectively.
First let us prove \ (8). To obtain the asymptotic formulas for large
eigenvalues we introduce a large parameter $\rho.$ If the considered
eigenvalue is of order $\rho^{2}$ we write the potential $q(x)\in W_{2}%
^{s}(F)$ in the form
\begin{equation}
q(x)=\sum_{\gamma\in\Gamma(\rho^{\alpha})}q_{\gamma}e^{i(\gamma,x)}%
+O(\rho^{-p\alpha}),
\end{equation}
where $p=s-d,$ $\Gamma(\rho^{\alpha})=\{\gamma\in\Gamma:0<$ $\mid\gamma
\mid<\rho^{\alpha})\}$ and $\alpha$ is defined in (7). Note that the relation
$q(x)\in W_{2}^{s}(F)$ means that
\[
\sum_{\gamma\in\Gamma}\mid q_{\gamma}\mid^{2}(1+\mid\gamma\mid^{2s})<\infty.
\]
This implies that if $s\geq d,$ then
\begin{equation}
\sum_{\gamma\in\Gamma}\mid q_{\gamma}\mid<c_{1},\text{ }\sup\mid\sum
_{\gamma\notin\Gamma(\rho^{\alpha})}q_{\gamma}e^{i(\gamma,x)}\mid\leq
\sum_{\mid\gamma\mid\geq\rho^{\alpha}}\mid q_{\gamma}\mid=O(\rho^{-p\alpha}),
\end{equation}
i.e., (21) holds. Here and in subsequent estimations we denote by $c_{i}$
($i=1,2,...)$ the positive, independent of $\rho,$ constants.

In [4] ( see Theorem 3.1, Theorem 6.1, and (6.45) of [4]) we proved that if
\begin{equation}
j\in S_{1}(\rho),\text{ }\beta\in S_{2}(\rho),\text{ }v\in S_{3}(\beta
,\rho),\tau\in S_{4}(\beta,j,v,\rho),
\end{equation}
then there exists unique simple eigenvalue $\Lambda_{N}(t),$ denoted in [4] by
$\Lambda(\lambda_{j,\beta})$ and in this paper by $\Lambda_{j,\beta}(v,\tau)$
or by $\Lambda_{N(j,\beta)}(t),$ of $L_{t}(q)$ satisfying
\begin{equation}
\Lambda_{j,\beta}(v,\tau)=\lambda_{j,\beta}(v,\tau)+O(\rho^{-\alpha_{2}})
\end{equation}
and the corresponding eigenfunction $\Psi_{N,t}(x),$ denoted here by
$\Psi_{j,\beta}(x),$ satisfies
\begin{equation}
\Psi_{j,\beta}(x)=\Phi_{j,\beta}(x)+O(\rho^{-\alpha_{2}}\ln\rho),
\end{equation}
where $\alpha_{2}$ is defined in (7) and the sets $S_{k}$ for $k=1,2,3,4$ are
defined as follows:
\begin{align}
S_{1}(\rho)  &  =\{j\in\mathbb{Z}:\mid j\mid<\frac{\rho^{\alpha_{1}}}%
{2\mid\delta\mid^{2}}-\frac{3}{2}\},\\
S_{2}(\rho)  &  =\{\beta\in\Gamma_{\delta}:\beta\in(R_{\delta}(\frac{3}{2}%
\rho-d_{\delta}-1)\backslash R_{\delta}(\frac{1}{2}\rho+d_{\delta
}+1))\backslash(\bigcup_{b\in\Gamma_{\delta}(\rho^{\alpha_{d}})}V_{b}^{\delta
}(\rho^{\frac{1}{2}}))\},\nonumber
\end{align}

where $d_{\delta}=\sup_{x,y\in F_{\delta}}\mid x-y\mid$ is the diameter of
$F_{\delta},$%
\[
R_{\delta}(c)=\{x\in H_{\delta}:\mid x\mid<c\},\ \Gamma_{\delta}%
(c)=\{b\in\Gamma_{\delta}:0<\mid b\mid<c\},
\]%
\[
V_{b}^{\delta}(c)=\{x\in H_{\delta}:\mid\mid x+b\mid^{2}-\mid x\mid^{2}%
\mid<c\},
\]%
\begin{equation}
S_{3}(\beta,\rho)=W(\rho)\backslash A(\beta,\rho),
\end{equation}
where $W(\rho)\equiv\{v\in(0,1):\mid\mu_{j}(v)-\mu_{j^{^{\prime}}}%
(v)\mid>\frac{2}{\ln\rho},$ $\forall j^{^{\prime}},j\in\mathbb{Z},$
$j^{^{\prime}}\neq j\},$%
\[
A(\beta,\rho)=\bigcup_{b\in\Gamma_{\delta}(\rho^{\alpha_{d}})}A(\beta
,b,\rho),
\]%
\[
A(\beta,b,\rho)=\{v\in\lbrack0,1):\exists j\in\mathbb{Z},\mid2(\beta,b)+\mid
b\mid^{2}+\mid(j+v)\delta\mid^{2}\mid<4d_{\delta}\rho^{\alpha_{d}}\},
\]
and$\ S_{4}(\beta,j,v,\rho)$ is an asymptotically full subset of $F_{\delta}%
$:
\begin{equation}
\text{ }\mu(S_{4}(\beta,j,v,\rho))=\mu(F_{\delta})(1+O(\rho^{-\alpha})).
\end{equation}

In this paper to obtain the asymptotic formulas, which is suitable for the
constructively determination of the spectral invariants, we put an additional
conditions on $\beta,$ namely we suppose that
\begin{equation}
\beta\notin(\bigcup_{b\in\Gamma_{\delta}(p\rho^{\alpha})}V_{b}^{\delta}%
(\rho^{a}))),
\end{equation}
where $a$ is defined in (9). By definition of $V_{b}^{\delta}(\rho^{a})$ the
relation (29) yields
\begin{equation}
\mid\mid\beta\mid^{2}-\mid\beta+\beta_{1}\mid^{2}\mid\geq\rho^{a},\text{
}\forall\beta_{1}\in\Gamma_{\delta}(p\rho^{\alpha}).
\end{equation}
Using the inequalities $\mid\beta_{1}\mid<p\rho^{\alpha},$ $\mid\tau
\mid<d_{\delta},$ $a>2\alpha,$ we obtain%
\begin{equation}
\mid\mid\beta+\tau\mid^{2}-\mid\beta+\beta_{1}+\tau\mid^{2}\mid>\frac{8}%
{9}\rho^{a},\text{ }\forall\beta_{1}\in\Gamma_{\delta}(p\rho^{\alpha}).
\end{equation}

Now we prove (8) by using (23), (31), and the following relation called the
binding formula for $L_{t}(q)$ and $L_{t}(q^{\delta})$:
\begin{equation}
(\Lambda_{N}(t)-\lambda_{j,\beta})b(N,j,\beta)=(\Psi_{N,t}(x),(q(x)-q^{\delta
}(x))\Phi_{j,\beta}(x)),
\end{equation}
where $b(N,j,\beta)=(\Psi_{N,t}(x),\Phi_{j,\beta}(x)),$ which can be obtained
from
\[
L_{t}(q)\Psi_{n,t}(x)=\Lambda_{n}(t)\Psi_{n,t}(x)
\]
by multiplying by $\Phi_{j,\beta}(x)$ and using $L_{t}(q^{\delta}%
)\Phi_{j,\beta}(x)=\lambda_{j,\beta}\Phi_{j,\beta}(x).$ In [4], using (21), we
proved that ( see (3.22) and (3.23) of [4]) if \ $\mid j\delta\mid<r,$
$\mid\beta\mid>\frac{1}{2}\rho,$ where $r\geq r_{1}$ and $r_{1}=\frac
{\rho^{\alpha_{1}}}{2\mid\delta\mid}+2\mid\delta\mid$, then the following
decomposition%
\begin{equation}
(q(x)-q^{\delta}(x))\Phi_{j,\beta}(x)=\sum\limits_{(j_{1},\beta_{1})\in
Q(\rho^{\alpha},9r)}A(j,\beta,j+j_{1,}\beta+\beta_{1})\Phi_{j+j_{1,}%
\beta+\beta_{1}}(x)+O(\rho^{-p\alpha})
\end{equation}
of $(q(x)-q^{\delta}(x))\Phi_{j,\beta}(x)$ by eigenfunction of $L_{t}%
(q^{\delta})$ holds, where

$Q(\rho^{\alpha},9r)=\{(j,\beta):\mid j\delta\mid<9r,$ $0<\mid\beta\mid
<\rho^{\alpha}\}$ and%
\begin{equation}
\sum\limits_{(j_{1},\beta_{1})\in Q(\rho^{\alpha},9r)}\mid A(j,\beta
,j+j_{1,}\beta+\beta_{1})\mid<c_{2}.
\end{equation}
Using this decomposition in (32), we get
\[
(\Lambda_{N}(t)-\lambda_{j,\beta})b(N,j,\beta)=O(\rho^{-p\alpha})
\]

\begin{equation}
+\sum\limits_{(j_{1},\beta_{1})\in Q(\rho^{\alpha},9r)}A(j,\beta,j+j_{1,}%
\beta+\beta_{1})b(N,j+j_{1},\beta+\beta_{1}).
\end{equation}

\begin{remark}
If $\mid j^{^{\prime}}\delta\mid<r,$ $\mid\beta^{^{\prime}}\mid>\frac{1}%
{2}\rho$ and $\mid\Lambda_{N}-\lambda_{j^{^{\prime}},\beta^{^{\prime}}}%
\mid>c(\rho),$ then by (35) we have
\[
b(N,j^{^{\prime}},\beta^{^{\prime}})=%
%TCIMACRO{\dsum _{(j_{1},\beta_{1})\in Q(\rho^{\alpha},9r)}}%
%BeginExpansion
{\displaystyle\sum_{(j_{1},\beta_{1})\in Q(\rho^{\alpha},9r)}}
%EndExpansion
\dfrac{A(j^{^{\prime}},\beta^{^{\prime}},j^{^{\prime}}+j_{1},\beta^{^{\prime}%
}+\beta_{1})b(N,j^{^{\prime}}+j_{1},\beta^{^{\prime}}+\beta_{1})}{\Lambda
_{N}-\lambda_{j^{^{\prime}},\beta^{^{\prime}}}}+O(\frac{1}{\rho^{p\alpha
}c(\rho)}).
\]
If $j\in S_{1}(\rho)$ then $\mid j\delta\mid<r_{1}=O(\rho^{\alpha_{1}})$ and
in (35) instead of $r$ we take $r_{1}.$
\end{remark}

\begin{theorem}
If (23) and (29) hold then the eigenvalue $\Lambda_{N(j,\beta)}(t)\equiv
\Lambda_{j,\beta}(v,\tau)$ defined in (24) satisfies (8).
\end{theorem}

\begin{proof}
Since $b(N,j,\beta)=1+O(\rho^{-\alpha_{2}}\ln\rho)$ ( see (25)), where
$N=N(j,\beta)$, we need to prove that the right-hand side of (35) is
$O(\rho^{-a}).$ First we show that
\begin{equation}
b(N,j+j_{1},\beta+\beta_{1})=O(\rho^{-a})
\end{equation}
for $\beta_{1}\in\Gamma_{\delta}(p\rho^{\alpha}),$ $j=o(\rho^{\frac{a}{2}}),$
$j_{1}=o(\rho^{\frac{a}{2}}).$ For this we prove the inequality
\begin{equation}
\mid\Lambda_{N}(t)-\lambda_{j+j_{1},\beta+\beta_{1}}\mid>\frac{1}{2}\rho
^{a},\text{ }\forall\beta_{1}\in\Gamma_{\delta}(p\rho^{\alpha}),\text{
}\forall j=o(\rho^{\frac{a}{2}}),\forall j_{1}=o(\rho^{\frac{a}{2}})
\end{equation}
and use the formula
\begin{equation}
b(N,j+j_{1},\beta_{1}+\beta)=\frac{(\Psi_{N,t}(x),(q(x)-q^{\delta}%
(x))\Phi_{j+j_{1},\beta_{1}+\beta}(x))}{\Lambda_{N}-\lambda_{j+j_{1},\beta
_{1}+\beta}}%
\end{equation}
which can be obtained from (32) by replacing the indices $j,\beta$ with
$j+j_{1},\beta+\beta_{1}.$ By (24) the inequality (37) holds if
\[
\mid\mu_{j}(v)+\mid\beta+\tau\mid^{2}-\mu_{j+j_{1}}(v)-\mid\beta+\beta
_{1}+\tau\mid^{2}\mid>\frac{5}{9}\rho^{a}.
\]
This inequality is consequence of inequality (31) and the equalities

$j=o(\rho^{\frac{a}{2}}),$ $j+j_{1}=o(\rho^{\frac{a}{2}})$ ( see the
conditions on $j,$ $j_{1}$ in (36), (37)),
\begin{equation}
\mu_{n}(v)=\mid(n+v)\delta\mid^{2}+O(\frac{1}{n}).
\end{equation}
Thus (36) is proved. Using (36), the definition of \ $Q(\rho^{\alpha}%
,9r_{1}),$ and the relations $r_{1}=O(\rho^{\alpha_{1}})$ ( see Remark 1),
$\alpha_{1}<\frac{a}{2}$ ( see (7), (9)) we obtain that all multiplicands
$b(N,j+j_{1},\beta+\beta_{1})$ in the right-hand side of (35), in the case
$r=r_{1},$ is $O(\rho^{-a}).$ Therefore (34) implies that the right-hand side
of (35) is $O(\rho^{-a})$
\end{proof}

To prove the asymptotic formula (9), we iterate (35) , in the case $r=r_{1},$
as follows. If $\mid j\delta\mid<r_{1},$ then the summation in (35) is taken
under condition

$(j_{1},\beta_{1})\in Q(\rho^{\alpha},9r_{1})$\ ( see Remark 1). By the
definition of $Q(\rho^{\alpha},9r_{1})$ we have $\mid j_{1}\delta\mid<9r_{1}.$
Hence $\mid(j+j_{1})\delta\mid<r_{2},$ where $r_{2}=10r_{1}.$ Therefore, using
(37) and Remark 1, we get
\[
b(N,j+j_{1},\beta_{1}+\beta)=%
%TCIMACRO{\dsum _{(j_{2},\beta_{2})\in Q(\rho^{\alpha},9r_{2})}}%
%BeginExpansion
{\displaystyle\sum_{(j_{2},\beta_{2})\in Q(\rho^{\alpha},9r_{2})}}
%EndExpansion
\dfrac{A(j(1),\beta(1),j(2),\beta(2))b(N,j(2),\beta(2))}{\Lambda_{N}%
-\lambda_{j+j_{1},\beta+\beta_{1}}},
\]
where $j(k)=j+j_{1}+j_{2}+...+j_{k},$ $\beta(k)=\beta+\beta_{1}+\beta
_{2}+...+\beta_{k}$ for $k=0,1,2,....$Using this in (35) we obtain
\begin{equation}
(\Lambda_{N}-\lambda_{j,\beta})b(N,j,\beta)=O(\rho^{-p\alpha})+
\end{equation}%
\[%
%TCIMACRO{\dsum _{\substack{(j_{1},\beta_{1})\in Q(\rho^{\alpha},9r_{1}%
%)\\(j_{2},\beta_{2})\in Q(\rho^{\alpha},9r_{2})}}}%
%BeginExpansion
{\displaystyle\sum_{\substack{(j_{1},\beta_{1})\in Q(\rho^{\alpha}%
,9r_{1})\\(j_{2},\beta_{2})\in Q(\rho^{\alpha},9r_{2})}}}
%EndExpansion
\dfrac{A(j,\beta,j(1),\beta(1))A(j(1),\beta(1),j(2),\beta(2))b(N,j(2),\beta
(2))}{\Lambda_{N}-\lambda_{j+j_{1},\beta+\beta_{1}}}.
\]
To prove (9) we use this formula and the following lemma.

\begin{lemma}
Suppose (23) and (29) hold. If $j^{^{\prime}}\neq j,$ $\mid j^{^{\prime}%
}\delta\mid<r,$ where

$r=O(\rho^{\frac{1}{2}\alpha_{2}}),$ $r\geq r_{1},$ and $r_{1}=\frac
{\rho^{\alpha_{1}}}{2\mid\delta\mid}+2\mid\delta\mid$ then
\[
b(N(j,\beta),j^{^{\prime}},\beta)=O(\rho^{-2a}r^{2}\ln\rho).
\]

\end{lemma}

\begin{remark}
If (23) holds ,then there exists unique index $N(j,\beta,v,\tau)$ , depending
on $j,\beta,v,\tau,$ for which the eigenvalue$\Lambda_{N}(t)$ satisfies (8).
Instead of $N(j,\beta,v,\tau)$ we write $N(j,\beta)$ ( or $N$) if $v,\tau$ (
or $j,\beta,v,\tau$) are unambiguous.
\end{remark}

\begin{theorem}
If (23) and (29) hold, then $\Lambda_{j,\beta}(v,\tau)$ satisfies (9).
\end{theorem}

\begin{proof}
We prove this by using (40). To estimate the summation in the right side of
(40) we divide the terms in this summation into three group. First, second,
and third group terms are the terms with multiplicands $b(N,j,\beta),$
$b(N,j(2),\beta)$ with $j(2)\neq j,$ and $b(N,j(2),\beta(2))$ with
$\beta(2)\neq\beta$ respectively. The sum of the first group terms is
$C_{1}(\Lambda_{N})b(N,j,\beta),$ where
\begin{equation}
C_{1}(\Lambda_{N})=%
%TCIMACRO{\dsum _{(j_{1},\beta_{1})\in Q(\rho^{\alpha},9r_{1})}}%
%BeginExpansion
{\displaystyle\sum_{(j_{1},\beta_{1})\in Q(\rho^{\alpha},9r_{1})}}
%EndExpansion
\dfrac{A(j,\beta,j+j_{1,}\beta+\beta_{1})A(j+j_{1},\beta+\beta_{1},j,\beta
)}{\Lambda_{N}-\lambda_{j+j_{1},\beta+\beta_{1}}}.
\end{equation}
The sum of the second group terms is
\[%
%TCIMACRO{\dsum _{\substack{(j_{1},\beta_{1})\in Q(\rho^{\alpha},9r_{1}%
%)\\(j_{2},\beta_{2})\in Q(\rho^{\alpha},9r_{2})}}}%
%BeginExpansion
{\displaystyle\sum_{\substack{(j_{1},\beta_{1})\in Q(\rho^{\alpha}%
,9r_{1})\\(j_{2},\beta_{2})\in Q(\rho^{\alpha},9r_{2})}}}
%EndExpansion
\dfrac{A(j,\beta,j+j_{1,}\beta+\beta_{1})A(j+j_{1},\beta+\beta_{1}%
,j(2),\beta)}{\Lambda_{N}-\lambda_{j+j_{1},\beta+\beta_{1}}}b(N,j(2),\beta),
\]
where $j(2)\neq j.$ Since $r_{2}=10r_{1}=O(\rho^{\alpha_{1}})$ ( see Remark 1)
the conditions on $j,$ $j_{1},$ $j_{2}$ and Lemma 1 imply that $j(2)=O(\rho
^{\alpha_{1}})$ and $b(N,j(2),\beta)=O(\rho^{-2a+2\alpha_{1}}\ln\rho).$ Using
this, (34) and (37) we obtain that the sum of the second group terms is
$O(\rho^{-3a+2\alpha_{1}}\ln\rho).$ The sum of the third group terms is \
\begin{equation}%
%TCIMACRO{\dsum _{\substack{(j_{1},\beta_{1})\in Q(\rho^{\alpha},9r_{1}%
%)\\(j_{2},\beta_{2})\in Q(\rho^{\alpha},9r_{2})}}}%
%BeginExpansion
{\displaystyle\sum_{\substack{(j_{1},\beta_{1})\in Q(\rho^{\alpha}%
,9r_{1})\\(j_{2},\beta_{2})\in Q(\rho^{\alpha},9r_{2})}}}
%EndExpansion
\dfrac{A(j,\beta,j(1),\beta(1))A(j(1),\beta(1),j(2),\beta(2))}{\Lambda
_{N}-\lambda_{j+j_{1},\beta+\beta_{1}}}b(N,j(2),\beta(2)),
\end{equation}
where $\beta(2)\neq\beta.$ Using (37) and Remark 1 we get%
\[
b(N,j(2),\beta(2))=\sum_{(j_{3},\beta_{3})\in Q(\rho^{\alpha},9r_{3})}%
\dfrac{A(j(2),\beta(2),j(3),\beta(3))b(N,j(3),\beta(3))}{\Lambda_{N}%
-\lambda_{j(2),\beta(2)}}+O(\rho^{-p\alpha}),
\]
where $r_{3}=10r_{2}.$ Substituting it into (42) and isolating the terms with
multiplicands $b(N,j,\beta)$ we see that the sum of the third group terms is%
\[
C_{2}(\Lambda_{N})b(N,j,\beta)+C_{3}(\Lambda_{N})+O(\rho^{-p\alpha}),
\]
where%
\begin{equation}
C_{2}(\Lambda_{N})=\sum_{\substack{(j_{1},\beta_{1})\in Q(\rho^{\alpha}%
,9r_{1}),\\(j_{2},\beta_{2})\in Q(\rho^{\alpha},90r_{1})}}\dfrac
{A(j,\beta,j(1),\beta(1))A(j(1),\beta(1),j(2),\beta(2))A(j(2),\beta
(2),j,\beta)}{(\Lambda_{N}-\lambda_{j+j_{1},\beta+\beta_{1}})(\Lambda
_{N}-\lambda_{j(2),\beta(2)})},
\end{equation}
\
\[
C_{3}(\Lambda_{N})=%
%TCIMACRO{\dsum _{\substack{(j_{1},\beta_{1})\in Q(\rho^{\alpha},9r_{1}%
%)\\(j_{2},\beta_{2})\in Q(\rho^{\alpha},9r_{2}),\\(j_{3},\beta_{3})\in
%Q(\rho^{\alpha},9r_{3})}}}%
%BeginExpansion
{\displaystyle\sum_{\substack{(j_{1},\beta_{1})\in Q(\rho^{\alpha}%
,9r_{1})\\(j_{2},\beta_{2})\in Q(\rho^{\alpha},9r_{2}),\\(j_{3},\beta_{3})\in
Q(\rho^{\alpha},9r_{3})}}}
%EndExpansion
\dfrac{(%
%TCIMACRO{\dprod \limits_{k=1,2,3}}%
%BeginExpansion
{\displaystyle\prod\limits_{k=1,2,3}}
%EndExpansion
A(j(k-1),\beta(k-1),j(k),\beta(k)))b(N,j(3),\beta(3))}{(\Lambda_{N}%
-\lambda_{j(1),\beta(1)})(\Lambda_{N}-\lambda_{j(2),\beta(2)})},
\]
\ and $(j(3),\beta(3))\neq(j,\beta).$ By (36) and Lemma 1 $b(N,j(3),\beta
(3))=O(\rho^{-a})$ for $(j(3),\beta(3))\neq(j,\beta).$ Using this, (34), and
taking into account that
\[
\mid\Lambda_{N}(t)-\lambda_{j(1),\beta(1)}\mid>\frac{1}{3}\rho^{a},\text{
}\mid\Lambda_{N}(t)-\lambda_{j(2),\beta(2)}\mid>\frac{1}{3}\rho^{a}%
\]
for $\beta(1)\neq\beta$ , $\beta(2)\neq\beta$ ( see (37)), we obtain
$C_{3}(\Lambda_{N})=O(\rho^{-3a}).$ The estimations of the first, second and
third groups terms imply that the formula (40) can be written in the form
\begin{equation}
(\Lambda_{N}-\lambda_{j,\beta})b(N,j,\beta)=(C_{1}(\Lambda_{N})+C_{2}%
(\Lambda_{N}))b(N,j,\beta)+O(\rho^{-3a+2\alpha_{1}}\ln\rho),
\end{equation}
where $N=N(j,\beta,v,\tau),$ $\Lambda_{N}=\Lambda_{j,\beta}(v,\tau).$
Therefore dividing both part of (44) by $b(N,j,\beta),$ where $b(N,j,\beta
)=1+o(1)$ (see (25)), we get
\begin{equation}
\Lambda_{j,\beta}=\lambda_{j,\beta}+C_{1}(\Lambda_{j,\beta}))+C_{2}%
(\Lambda_{j,\beta}))+O(\rho^{-3a+2\alpha_{1}}\ln\rho).
\end{equation}
The calculations in Appendix C and in Appendix B show that
\begin{equation}
C_{1}(\Lambda_{j,\beta}(v,\tau))=\frac{1}{4}\int_{F}\left\vert f_{\delta
,\beta+\tau}(x)\right\vert ^{2}\left\vert \varphi_{j,v}\right\vert
^{2}dx+O(\rho^{-3a+2\alpha_{1}}),
\end{equation}%
\begin{equation}
C_{2}(\Lambda_{j,\beta}(v,\tau))=O(\rho^{-3a+2\alpha_{1}}).
\end{equation}
Therefore (9) follows from (45)
\end{proof}

\begin{theorem}
If \ (23) and (29) hold then the eigenfunction $\Psi_{j,\beta}(x)$
corresponding to $\Lambda_{j,\beta}(v,\tau)$ satisfies (10).
\end{theorem}

\begin{proof}
To prove (10) we need to show that
\begin{equation}
\sum_{(j^{^{\prime}},\beta^{^{\prime}}):(j^{^{\prime}},\beta^{^{\prime}}%
)\neq(j,\beta)}\mid b(N(j,\beta),j^{^{\prime}},\beta^{^{\prime}})\mid
^{2}=O(\rho^{-2a}).
\end{equation}
In [4] ( see (6.36) of [4]) we proved that
\begin{equation}
\sum_{(j^{^{\prime}},\beta^{^{\prime}})\in S^{c}(k-1)}\mid b(N,j^{^{\prime}%
},\beta^{^{\prime}})\mid^{2}=O(\rho^{-2k\alpha_{2}}(\ln\rho)^{2}),
\end{equation}
where $S^{c}(n)=K_{0}\backslash S(n),$ $K_{0}=\{(j^{^{\prime}},\beta
^{^{\prime}}):j^{^{\prime}}\in\mathbb{Z},\beta^{^{\prime}}\in\Gamma_{\delta
},(j^{^{\prime}},\beta^{^{\prime}})\neq(j,\beta)\},$ $S(n)=\{(j^{^{\prime}%
},\beta^{^{\prime}})\in K_{0}:\mid\beta-\beta^{^{\prime}}\mid\leq
n\rho^{\alpha}$ $,\mid j^{^{\prime}}\delta\mid<10^{n}h\},$ $h=O(\rho^{\frac
{1}{2}\alpha_{2}})$ and $k$ can be chosen such that $k\alpha_{2}>a,$ $k<p.$
Therefore it is enough to prove that
\begin{equation}
\sum_{(j^{^{\prime}},\beta^{^{\prime}})\in S(k-1)}\mid b(N,j^{^{\prime}}%
,\beta^{^{\prime}})\mid^{2}=O(\rho^{-2a}).
\end{equation}
Using (37), (38), definition of $S(k-1)$ and Bessel inequality for the basis
$\{\Phi_{j^{^{\prime}},\beta^{^{\prime}}}(x):j^{^{\prime}}\in\mathbb{Z},$
$\beta^{^{\prime}}\in\Gamma_{\delta}\}$ we have
\begin{equation}
\sum_{(j^{^{\prime}},\beta^{^{\prime}}):(j^{^{\prime}},\beta^{^{\prime}})\in
S(k-1),\beta^{^{\prime}}\neq\beta}\mid b(N,j^{^{\prime}},\beta^{^{\prime}%
})\mid^{2}=\nonumber
\end{equation}%
\begin{equation}
\sum_{(j^{^{\prime}},\beta^{^{\prime}})}\frac{\mid(\Psi_{N}(x)(q(x)-Q(s)),\Phi
_{j^{^{\prime}},\beta^{^{\prime}}}(x))\mid^{2}}{\mid\Lambda_{N}-\lambda
_{j^{^{\prime}},\beta^{^{\prime}}}\mid^{2}}=O(\rho^{-2a}).
\end{equation}
In the case $\beta^{^{\prime}}=\beta,$ $j^{^{\prime}}\neq j$ using Lemma 1 (
we can use it since $\mid j^{^{\prime}}\delta\mid=O(\rho^{\frac{1}{2}%
\alpha_{2}})$ for $(j^{^{\prime}},\beta^{^{\prime}})\in S(k-1)),$ we obtain
\begin{equation}
\sum_{(j^{^{\prime}},\beta)\in S(k-1),j^{^{\prime}}\neq j}\mid b(N,j^{^{\prime
}},\beta)\mid^{2}=O(\rho^{-4a+2\alpha_{2}}(\ln\rho)^{2})K,
\end{equation}
where $K$ is the number of $j^{^{\prime}}$ satisfying $(j^{^{\prime}}%
,\beta)\in S(k-1).$ It is clear that $K=O(\rho^{\frac{1}{2}\alpha_{2}})$.
Since $\alpha_{2}<\frac{a}{2}$ ( see (7) and (9)), the right side of (52) is
$O(\rho^{-2a}).$ Therefore (52) and (51) give (50)
\end{proof}

Now we estimate the derivatives of $\Lambda_{N}(t)$ by using the following lemma.

\begin{lemma}
Let $\Lambda_{N}(\beta+\tau+(j+v)\delta),$ be a simple eigenvalue of $L_{t}$
satisfying
\begin{equation}
\mid\Lambda_{N}(\beta+\tau+(j+v)\delta)-\lambda_{j,\beta}(v,\tau)\mid<1,
\end{equation}
where $j,\beta,$ satisfy (23), and $\beta+\tau+(j+v)\delta-t\in\Gamma$. Then
\begin{equation}
\mid\beta+\tau\mid\frac{\partial\Lambda_{N}(t)}{\partial h}=\sum_{j^{^{\prime
}}\in\mathbb{Z},\text{ }\beta^{^{\prime}}\in\Gamma_{\delta}}(\beta+\tau
,\beta^{^{\prime}}+\tau)\mid b(N,j^{^{\prime}},\beta^{^{\prime}})\mid^{2},
\end{equation}
where $\frac{\partial\Lambda_{N}(t)}{\partial h}$ is the derivative of
$\Lambda_{N}(t)$ in the direction $h=\frac{\beta+\tau}{\mid\beta+\tau\mid}.$
Moreover
\begin{equation}
\mid b(N,j^{^{\prime}},\beta^{^{\prime}})\mid\leq\frac{c_{3}}{(\mid
\beta^{^{\prime}}+\tau\mid^{2}+\mid(j^{^{\prime}}+v)\delta\mid^{2})\mid
\beta^{^{\prime}}+\tau\mid^{2d+6}}%
\end{equation}
for all $\beta^{^{\prime}}$ $\ $satisfying $\mid\beta^{^{\prime}}+\tau\mid
\geq4\rho$ and for all $j^{^{\prime}}\in\mathbb{Z}$ .
\end{lemma}

\begin{theorem}
If (23) and (29) hold, then (11) holds too.
\end{theorem}

\begin{proof}
It follows from (55), (48), (10) that
\begin{align*}
\sum_{j^{^{\prime}}\in\mathbb{Z},\mid\beta^{^{\prime}}+\tau\mid\geq4\rho
}(\beta+\tau,\beta^{^{\prime}}+\tau)  &  \mid b(N,j^{^{\prime}},\beta
^{^{\prime}})\mid^{2}=O(\rho^{2-2a}),\\
\sum_{j^{^{\prime}}\in\mathbb{Z},\mid\beta^{^{\prime}}+\tau\mid<4\rho
,(j^{^{\prime}},\beta^{^{\prime}})\neq(j,\beta)}(\beta+\tau,\beta^{^{\prime}%
}+\tau)  &  \mid b(N,j^{^{\prime}},\beta^{^{\prime}})\mid^{2}=O(\rho
^{2-2a}),\\
(\beta+\tau,\beta+\tau)  &  \mid b(N,j,\beta)\mid^{2}=\mid\beta+\tau\mid
^{2}+O(\rho^{2-2a}),
\end{align*}
where $N=N(j,\beta,v,\tau)$ ( see Remark 2). Therefore (11) follows from (54)
\end{proof}

To prove the main results of this paper we need the following lemmas

\begin{lemma}
If $\Lambda_{N}(\beta+\tau+(j+v)\delta)$ is a simple eigenvalue of $L_{t}(q),$ where

$\beta+\tau+(j+v)\delta-t\in\Gamma,$ satisfying (53) and $N\neq N(j,\beta)$,
then
\[
\mid\beta+\tau\mid\frac{\partial\Lambda_{N}(t)}{\partial h}<\mid\beta+\tau
\mid^{2}-\frac{1}{4}\rho^{2\alpha_{d}}.
\]

\end{lemma}

The proof of this lemma is given in Section 3. Here we only note some reason
of this estimations. It follows from (48) that
\begin{equation}
\mid b(N,j,\beta)\mid^{2}=1+O(\rho^{-2a})\text{ for }N=N(j,\beta).
\end{equation}
Since $\parallel\Phi_{j,\beta}(x)\parallel=1,$ using Parseval's equality for
the orthonormal basis

$\{\Psi_{N}(x):N=1,2,...,\}$ and (56), we get%
\begin{equation}
\mid b(N,j,\beta)\mid^{2}=O(\rho^{-2a}),\text{ }\forall\text{ }N\neq
N(j,\beta).
\end{equation}
This with the long technical estimations of the other term of the series of
the right side of (54) implies the proof of the Lemma 3.

\begin{lemma}
Let $b$ be a maximal element of $\Gamma_{\delta}$ and $v\in(0,\frac{1}{2}%
)\cup(\frac{1}{2},1).$ Then there exist $\rho(v)$ such that for $\rho\geq
\rho(v)$ there exists $\beta\in S_{2}(\rho)$ satisfying the relation $v\notin
A(\beta,\rho)$ and the inequalities
\begin{align}
\frac{1}{3}|\rho|^{a}  &  <\mid(\beta+\tau,b)\mid<3|\rho|^{a},\\
\  &  \mid(\beta+\tau,\gamma)\mid>\frac{1}{3}|\rho|^{a},\text{ }\forall
\gamma\in S(\delta,b)\backslash\delta\mathbb{R}\text{,}\\
\  &  \mid(\beta+\tau,\gamma)\mid>\frac{1}{3}|\rho|^{a+2\alpha},\text{
}\forall\gamma\not \in S(\delta,b),\text{ }\mid\gamma\mid<|\rho|^{\alpha
}\text{,}%
\end{align}%
\begin{equation}
\int_{F}\left\vert f_{\delta,\beta+\tau}(x)\right\vert ^{2}\left\vert
\varphi_{n,v}\right\vert ^{2}dx<c_{4}\rho^{-2a}%
\end{equation}
for $\tau\in F_{\delta},$ where $S_{2},$ $A(\beta,\rho),$ $f_{\delta
,\beta+\tau},$ $S(\delta,b)$ are defined in (26), (27), (9), (13).
\end{lemma}

\begin{theorem}
Suppose $q(x)\in W_{2}^{s}(F)$ and the band functions are known. Then the
spectral invariants $\mu_{n}(v)$ for $n\in\mathbb{Z}$, $v\in\lbrack0,1)$ and
(12), (15), (16), (17), (20) can be constructively determined.
\end{theorem}

\begin{proof}
Take any $j\in\mathbb{Z}$ and $v\in(0,\frac{1}{2})\cup(\frac{1}{2},1).$ In [4]
( see Lemma 3.7 of [4]) we proved that
\[
(\varepsilon(\rho),\frac{1}{2}-\varepsilon(\rho))\cup(\frac{1}{2}%
+\varepsilon(\rho),1-\varepsilon(\rho))\subset W(\rho),
\]
where $W(\rho)$ is defined in (27) and $\varepsilon(\rho)\rightarrow0$ as
$\rho\rightarrow\infty.$ Therefore $v\in W(\rho)$ for $\rho\gg1.$ On the other
hand by Lemma 4 there exists $\beta\in S_{2}(\rho)$ such that $v\notin
A(\beta,\rho)$ and (58)-(61) holds. Then $v\in S_{3}(\beta,\rho)$ ( see (27)).
Thus $j,\beta,v$ satisfy (23) and $\beta$ satisfies (58)-(61) for $\rho\gg1.$
Replacing $\rho$ by $\rho_{k}\equiv3^{k}\rho$ for

$k=1,2,...,$ in the same way we obtain the sequence $\beta_{1},\beta_{2},$...,
such that $\beta_{k}\in S_{2}(\rho_{k}),$ $v\in S_{3}(\beta_{k},\rho_{k})$ and
the inequalities obtained from (58)-(61) by replacing $\beta,\rho$ with
$\beta_{k},\rho_{k}$ holds. Now take $\tau$ from $F_{\delta}$ and consider the
band functions $\Lambda_{N}(\beta_{k}+\tau+(j+v)\delta)$ for $N=1,2,...$. Let
$A_{k}$ be the set of all $\tau\in F_{\delta}$ for which there is $N$
satisfying the conditions:%
\begin{align}
&  \mid\Lambda_{N}(\beta_{k}+\tau+(j+v)\delta)-\mid\beta_{k}+\tau\mid^{2}%
-\mid(j+v)\delta\mid^{2}\mid<1,\\
&  \Lambda_{N}(\beta_{k}+\tau+(j+v)\delta)\text{ is a simple eigenvalue,}\\
&  \mid\mid\beta_{k}+\tau\mid\frac{\partial\Lambda_{N}(\beta_{k}%
+\tau+(j+v)\delta)}{\partial h}-\mid\beta_{k}+\tau\mid^{2}\mid<\rho
_{k}^{2-2a+\alpha},
\end{align}
where $h=\frac{\beta_{k}+\tau}{\mid\beta_{k}+\tau\mid}.$ For $\tau\in A_{k}$
take one of eigenvalues $\Lambda_{N}(\beta_{k}+\tau+(j+v)\delta)$ satisfying
(62)-(64) and calculate the integral
\[
J(A_{k})=\frac{1}{\mu(F_{\delta})}\int_{A_{k}}(\Lambda_{N}(\beta_{k}%
+\tau+(j+v)\delta)-\mid\beta_{k}+\tau\mid^{2})d\tau.
\]
We write these integral as sum of $J(S_{4}(\beta_{k},j,v,\rho_{k}))$ and
$J(A_{k}\backslash S_{4}(\beta_{k},j,v,\rho_{k})),$ where $S_{4}(\beta
_{k},j,v,\rho_{k})$ is defined in (28). Note that if $\tau\in S_{4}%
(\beta,j,v,\rho),$ where $j\in S_{1}(\rho),$ $\beta\in S_{2}(\rho),$ $v\in
S_{3}(\beta,\rho)$ ( see (23)) then the formulas (24), (8), (9), (10), (11)
hold. Here for brevity of notations instead of $S_{4}(\beta_{k},j,v,\rho_{k})$
we will write $S_{4}.$ If $\tau\in S_{4}$ then by (24) and Theorem 4 the
eigenvalue $\Lambda_{j,\beta_{k}}(v,\tau)$ satisfies the conditions (62)-(64)
and by Lemma 3 the other eigenvalues does not satisfy these conditions. Hence
in the integral $J(S_{4})$ instead of

$\Lambda_{N}(\beta_{k}+\tau+(j+v)\delta)$ we must take $\Lambda_{j,\beta_{k}%
}(v,\tau)).$ Therefore using (8), (28), the inclusion $A_{k}\subset F_{\delta
},$ and (62), we get
\[
J(S_{4})=\mu_{j}(v)+O(\rho_{k}^{-a}),\text{ }\mu(A_{k}\backslash S_{4}%
)=O(\rho_{k}^{-a}),\text{ }J(A_{k}\backslash S_{4})=O(\rho^{-a}).
\]
These equalities imply that
\[
J(A_{k})=\mu_{j}(v)+O(\rho_{k}^{-a}).
\]
Now tending $k$ to $\infty$ we find the eigenvalue $\mu_{j}(v)$ for
$j\in\mathbb{Z}$ and $v\in(0,\frac{1}{2})\cup(\frac{1}{2},1)$. Since $\mu
_{j}(0)$ and $\mu_{j}(\frac{1}{2})$ are the end points of the interval
$\{\mu_{j}(v):v\in(0,\frac{1}{2})\}$ the invariants $\mu_{j}(v)$ are
constructively determined for all $j\in\mathbb{Z}$, $v\in\lbrack0,1).$ In the
Appendix D we constructively determine (17) from the asymptotic formulas for
$\mu_{j}(v).$

Let $B_{k}$ be the set of $\tau\in F_{\delta}$ for which there is $N$
satisfying (63), (64), and
\begin{equation}
\mid\Lambda_{N}(\beta_{k}+\tau+(j+v)\delta)-\mid\beta_{k}+\tau\mid^{2}-\mu
_{j}(v)\mid<\rho_{k}^{-2a+\frac{\alpha}{2}},
\end{equation}
For $\tau\in B_{k}$ take one of the eigenvalues $\Lambda_{N}(\beta_{k}%
+\tau+(j+v)\delta)$ satisfying (63)-(65) and estimate the integral
\[
J^{^{\prime}}(B_{k})=\frac{\mid(\beta_{k}+\tau,b)\mid^{2}}{\mu(F_{\delta
})|b|^{4}}\int_{B_{k}}(\Lambda_{N}(\beta_{k}+\tau+(j+v)\delta)-\mid\beta
_{k}+\tau\mid^{2}-\mu_{j}(v))d\tau.
\]
We write these integral as sum of $J^{^{\prime}}(S_{4})$ and $J^{^{\prime}%
}(B_{k}\backslash S_{4}).$ If $\tau\in S_{4}$ then by Theorem 4 and Lemma 3
only $\Lambda_{j,\beta_{k}}(v,\tau))$ satisfies (63)-(65). Hence in the
integral $J^{^{\prime}}(S_{4})$ instead of $\Lambda_{N}(\beta_{k}%
+\tau+(j+v)\delta)$ we must take $\Lambda_{j,\beta_{k}}(v,\tau)).$ Therefore
using (9), (58) we get
\begin{equation}
J^{^{\prime}}(S_{4})=\frac{\mid(\beta_{k}+\tau,b)\mid^{2}}{\mu(F_{\delta
})|b|^{4}}\int_{S_{4}}\int_{F}\mid f_{\delta,\beta_{k}+\tau}(x)\varphi
_{j,v}\mid^{2}dxd\tau+O(\rho_{k}^{2\alpha_{1}-a}\ln\rho).
\end{equation}
Moreover using (65), (58), and $\mu(B_{k}\backslash S_{4})=O(\rho_{k}%
^{-\alpha})$ ( see (28)), we obtain
\begin{equation}
J^{^{\prime}}(B_{k}\backslash S_{4})=O(\rho_{k}^{-\frac{\alpha}{2}}).
\end{equation}
Substituting the decomposition $|\delta|^{-2}(\gamma,\delta)\delta
+|b|^{-2}(\gamma,b)b$ of $\gamma$ for $\gamma\in S(\delta,b),$ $\mid\gamma
\mid<|\rho_{k}|^{\alpha}$ into the denominator of the fraction in
$f_{\delta,\beta_{k}+\tau}(x)$ ( for definition of this function see (9)) and
using (58), (60) we have
\begin{equation}
\lim_{k\rightarrow\infty}|b|^{-2}(\beta_{k}+\tau,b)f_{\delta,\beta_{k}+\tau
}(x)=\sum_{\gamma\in S(\delta,b)\backslash\delta\mathbb{R}}\frac{\gamma
}{(\gamma,b)}q_{\gamma}e^{(\gamma,x)}\equiv q_{\delta,b}(x),
\end{equation}
where $q_{\delta,b}(x)$ is defined in (13) and the convergence of the series
(13) is proved in the proof of Lemma 4. This with (66) and (67) implies that
\begin{equation}
\lim_{k\rightarrow\infty}J^{^{\prime}}(B_{k})=\int_{F}\left\vert q_{\delta
,b}(x)\right\vert ^{2}\left\vert \varphi_{j,v}\right\vert ^{2}dx\equiv
J(\delta,b,j,v)
\end{equation}
( see (12)). In (69) tending $j$ to infinity and using (14) we get the
invariant $J_{0}(\delta,b)$ ( see (15)). Then we find the other invariants
$J_{1}(\delta,b),J_{2}(\delta,b),...,$ of (15) as follows

$J_{1}=\lim_{j\rightarrow\infty}(J-J_{0})j,$ $J_{2}=\lim_{j\rightarrow\infty
}((J-J_{0})j^{2}-J_{1}j),....$

In the Appendix D using the asymptotic formulas for the eigenfunctions of
$T_{v}(Q)$ we constructively determine the invariants (16), (20) from (15) and (17)
\end{proof}

\section{The proofs of the lemmas}

\ {\Large The proof of Lemma 1}

To prove this lemma we use the following formula obtained from (40) by
replacing $j$ and $r_{1}$ with $j^{^{\prime}}$ and $r$ respectively%
\begin{equation}
(\Lambda_{N(j,\beta)}-\lambda_{j^{^{\prime}},\beta})b(N,j^{^{\prime}}%
,\beta)=O(\rho^{-p\alpha})+
\end{equation}%
\[%
%TCIMACRO{\dsum _{\substack{(j_{1},\beta_{1})\in Q(\rho^{\alpha},9r)\\(j_{2}%
%,\beta_{2})\in Q(\rho^{\alpha},90r)}}}%
%BeginExpansion
{\displaystyle\sum_{\substack{(j_{1},\beta_{1})\in Q(\rho^{\alpha}%
,9r)\\(j_{2},\beta_{2})\in Q(\rho^{\alpha},90r)}}}
%EndExpansion
\dfrac{A(j,\beta,j^{^{\prime}}(1)_{,}\beta(1))A(j^{^{\prime}}(1)_{,}%
\beta(1),j^{^{\prime}}(2)_{,}\beta(2))b(N,j^{^{\prime}}(2),\beta(2))}%
{\Lambda_{N}-\lambda_{j^{^{\prime}}+j_{1},\beta+\beta_{1}}},
\]
where $j^{^{\prime}}(k)=j^{^{\prime}}+j_{1}+j_{2}+...+j_{k}$ for
$k=0,1,2,....$By (36) we have
\begin{equation}
b(N,j^{^{\prime}}(2),\beta(2))=O(\rho^{-a})
\end{equation}
for $\beta(2)\neq\beta.$ If $j^{^{\prime}}(2)\neq j,$ then using (8) and
taking into account that

$v\in S_{3}(\beta,\rho)\subset W(\rho)$ ( see the definition of $W(\rho)$ in
(27)) we obtain
\begin{equation}
\mid\Lambda_{N(j,\beta)}-\lambda_{j^{^{\prime}},\beta}\mid>\frac{1}{\ln\rho}.
\end{equation}
Therefore using (34), Remark 1, and (36) we see that
\[
b(N,j^{^{\prime}}(2),\beta)=O(\rho^{-a}\ln\rho)
\]
for $j^{^{\prime}}(2)\neq j.$ Using this, (34), and the estimations\ (37),
(71) we see that the sum of the terms of the right-hand side of (70) with
multiplicand $b(N,j^{^{\prime}}(2),\beta(2))$ for $(j^{^{\prime}}%
(2),\beta(2))$ $\neq(j,\beta)$ is $O(\rho^{-2a}\ln\rho).$ It means that the
formula (70) can be written in the form
\begin{equation}
(\Lambda_{N}-\lambda_{j^{^{\prime}},\beta})b(N,j^{^{\prime}},\beta
)=O(\rho^{-2a}\ln\rho)+C_{1}(j^{^{\prime}},\Lambda_{N})b(N,j,\beta),
\end{equation}
where%
\begin{equation}
C_{1}(j^{^{\prime}},\Lambda_{N})=\sum_{(j_{1},\beta_{1})\in Q(\rho^{\alpha
},9r)}\dfrac{A(j^{^{\prime}},\beta,j^{^{\prime}}+j_{1},\beta+\beta
_{1})A(j^{^{\prime}}+j_{1},\beta+\beta_{1},j,\beta)}{\Lambda_{N}%
-\lambda_{j^{^{\prime}}+j_{1},\beta+\beta_{1}}}.
\end{equation}
By (8), (37), and (34) we have%
\[
\frac{1}{\Lambda_{N}-\lambda_{j^{^{\prime}}+j_{1},\beta+\beta_{1}}}=\frac
{1}{\lambda_{j,\beta}-\lambda_{j^{^{\prime}}+j_{1},\beta+\beta_{1}}}%
=O(\rho^{-3a})
\]
and
\begin{equation}
C_{1}(j^{^{\prime}},\Lambda_{N})=C_{1}(j^{^{\prime}},\lambda_{j,\beta}%
)+O(\rho^{-3a}),
\end{equation}
where $C_{1}(j^{^{\prime}},\lambda_{j,\beta})$ is obtained from $C_{1}%
(j^{^{\prime}},\Lambda_{N})$ by replacing $\Lambda_{N}$ with $\lambda
_{j,\beta}$ in the denominator of the fractions in (74). In Appendix A we
prove that
\begin{equation}
C_{1}(j^{^{\prime}},\lambda_{j,\beta})=O(\rho^{-2a}r^{2})
\end{equation}
for $\mid j^{^{\prime}}\delta\mid<r,$ $(j_{1},\beta_{1})\in Q(\rho^{\alpha
},9r),$ $j\in S_{1}.$ Therefore dividing both sides of (73) by $\Lambda
_{N}-\lambda_{j^{^{\prime}},\beta}$ and using (72), (75), (76) we get the
proof of the lemma.

{\Large The proof of Lemma 2}

We calculate the derivative of $\Lambda_{N}(t)$ by using the formula
\[
\frac{\partial\Lambda_{N}(t)}{\partial t_{j}}=2t_{j}-2i(\frac{\partial
}{\partial x_{j}}\Phi_{N,t}(x),\Phi_{N,t}(x)),
\]
where $\Phi_{N,t}(x)=e^{-i(t,x)}\Psi_{N,t}(x),$ $t=(t_{1},t_{2},...,t_{d})$ (
see (5.12) of [4]). Then
\begin{equation}
\frac{\partial\Lambda_{N}(t)}{\partial h}=\sum_{j=1}^{d}h_{j}\frac
{\partial\Lambda_{N}(t)}{\partial t_{j}}=2(h,t)-2i(\frac{\partial}{\partial
h}\Phi_{N,t}(x),\Phi_{N,t}(x)).
\end{equation}
To compute $\frac{\partial}{\partial h}\Phi_{N,t}(x)$ we prove that the
decomposition
\begin{equation}
\Phi_{N,t}(x)=\sum_{j^{^{\prime}}\in\mathbb{Z},\beta^{^{\prime}}\in
\Gamma_{\delta}}b(N,j^{^{\prime}},\beta^{^{\prime}})e^{i(\beta^{^{\prime}%
}+\tau-t,x)}\varphi_{j^{^{\prime}}}((\delta,x))
\end{equation}
of $\Psi_{N,t}(x)$ over basis $\{\Psi_{j,\beta}(x):j\in\mathbb{Z},\beta
\in\Gamma_{\delta}\}$ can be differentiated term by term. Since $(\delta
,h)=0,$
\[
\frac{\partial}{\partial h}e^{i(\beta^{^{\prime}}+\tau-t,x)}\varphi
_{j^{^{\prime}}}((\delta,x))=i(\beta^{^{\prime}}+\tau-t,h)e^{i(\beta
^{^{\prime}}+\tau-t,x)}\varphi_{j^{^{\prime}}}((\delta,x)),
\]
we need to prove that%
\begin{equation}
\frac{\partial}{\partial h}\Phi_{N,t}(x)=\sum_{j^{^{\prime}}\in\mathbb{Z}%
,\beta^{^{\prime}}\in\Gamma_{\delta}}i(\beta^{^{\prime}}+\tau
-t,h)b(N,j^{^{\prime}},\beta^{^{\prime}})e^{i(\beta^{^{\prime}}+\tau
-t,x)}\varphi_{j^{^{\prime}}}((\delta,x)).
\end{equation}
Therefore we consider the convergence of these series by estimating the
multiplicand $b(N,j^{^{\prime}},\beta^{^{\prime}}).$ First we estimate this
multiplicand for $(j^{^{\prime}},\beta^{^{\prime}})\in E,$ where
$E=\{(j^{^{\prime}},\beta^{^{\prime}}):\mid(j^{^{\prime}}+v)\delta\mid
^{2}+\mid\beta^{^{\prime}}+\tau\mid^{2}\geq9\rho^{2}\}$ by using the formula
\begin{equation}
b(N,j^{^{\prime}},\beta^{^{\prime}})=\frac{(\Psi_{N,t}(x),(q(x)-q^{\delta
}(x))\Phi_{j^{^{\prime}},\beta^{^{\prime}}}(x))}{\Lambda_{N}-\lambda
_{j^{^{\prime}},\beta^{^{\prime}}}}%
\end{equation}
which can be obtained from (38) by replacing the indices $j+j_{1},\beta
+\beta_{1}$ with $j^{^{\prime}},\beta^{^{\prime}}.$ It follows from (8) and
(23) that%
\begin{equation}
\mid\Lambda_{N}\mid<3\rho^{2}.
\end{equation}
This inequality, the condition $(j^{^{\prime}},\beta^{^{\prime}})\in E,$
definition of $\lambda_{j^{^{\prime}},\beta^{^{\prime}}}$ and (39) give%
\begin{equation}
\lambda_{j^{^{\prime}},\beta^{^{\prime}}}-\Lambda_{N}>\frac{1}{2}%
(\mid(j^{^{\prime}}+v)\delta\mid^{2}+\mid\beta^{^{\prime}}+\tau\mid^{2}%
)>\rho^{2}%
\end{equation}
for $(j^{^{\prime}},\beta^{^{\prime}})\in E.$ Therefore (80) implies that
\begin{equation}
\mid b(N,j^{^{\prime}},\beta^{^{\prime}})\mid\leq\frac{c_{5}}{\mid
(j^{^{\prime}}+v)\delta\mid^{2}+\mid\beta^{^{\prime}}+\tau\mid^{2}},\text{
}\forall(j^{^{\prime}},\beta^{^{\prime}})\in E.
\end{equation}
Now we obtain the high order estimation for $b(N,j^{^{\prime}},\beta
^{^{\prime}})$ when $\mid\beta^{^{\prime}}+\tau\mid\geq4\rho$. In this case to
estimate $b(N,j^{^{\prime}},\beta^{^{\prime}})$ we use the iterations of the
formula in Remark 1. To iterate this formula we use the following obvious
relations
\[
\mid\beta^{^{\prime}}+\tau-\beta_{1}-\beta_{2}-...-\beta_{k}\mid^{2}>\frac
{3}{4}\mid\beta^{^{\prime}}+\tau\mid^{2}%
\]
for $k=1,2,...,d+3,$ where $\mid\beta_{i}\mid<\rho^{\alpha}$ for
$i=0,1,...,k.$ This and (81) give%

\begin{equation}
\lambda_{j^{^{\prime}}(k),\beta^{^{\prime}}(k)}-\Lambda_{N}>\frac{1}{5}%
\mid\beta^{^{\prime}}+\tau\mid^{2},\forall\mid\beta^{^{\prime}}+\tau\mid
\geq4\rho,
\end{equation}
where $\beta^{^{\prime}}(k)=\beta^{^{\prime}}+\beta_{1}+\beta_{2}%
+...+\beta_{k}.$ Moreover if $\mid j^{^{\prime}}\delta\mid<c,$ where $c$ is a
positive number, then $(j_{k},\beta_{k})\in Q(\rho^{\alpha},10^{k-1}9c).$
These conditions on $j^{^{\prime}}$ and $j_{1}$ imply that $\mid j^{^{\prime}%
}(1)\delta\mid<10c.$ Therefore in the formula in Remark 1 replacing
$j^{^{\prime}},\beta^{^{\prime}},r$ by $j^{^{\prime}}(1),\beta^{^{\prime}%
}(1),10c,$ we get
\[
b(N,j^{^{\prime}}(1),\beta^{^{\prime}}(1))=O(\rho^{-p\alpha})+
\]%
\[%
%TCIMACRO{\dsum _{(j_{2},\beta_{2})\in Q(\rho^{\alpha},90c)}}%
%BeginExpansion
{\displaystyle\sum_{(j_{2},\beta_{2})\in Q(\rho^{\alpha},90c)}}
%EndExpansion
\dfrac{A(j^{^{\prime}}(1),\beta^{^{\prime}}(1),j^{^{\prime}}(2),\beta
^{^{\prime}}(2))b(N,j^{^{\prime}}(2),\beta^{^{\prime}}(2))}{\Lambda
_{N}-\lambda_{j^{^{\prime}}(1),\beta^{^{\prime}}(1)}}.
\]
In the same way we obtain
\begin{equation}
b(N,j^{^{\prime}}(k),\beta^{^{\prime}}(k))=O(\rho^{-p\alpha})+
\end{equation}%
\[%
%TCIMACRO{\dsum _{(j_{k+1},\beta_{k+1})\in Q(\rho^{\alpha},(10^{k})9c)}}%
%BeginExpansion
{\displaystyle\sum_{(j_{k+1},\beta_{k+1})\in Q(\rho^{\alpha},(10^{k})9c)}}
%EndExpansion
\dfrac{A(j^{^{\prime}}(k),\beta^{^{\prime}}(k),j^{^{\prime}}(k+1),\beta
^{^{\prime}}(k+1))b(N,j^{^{\prime}}(k+1),\beta^{^{\prime}}(k+1))}{\Lambda
_{N}-\lambda_{j^{^{\prime}}(k),\beta^{^{\prime}}(k)}}%
\]
for $k=1,2,....$ In the formula in Remark 1 for $r=c$ using this formula for
$k=1,2,...d+3$ successively, we get%
\begin{equation}
b(N,j^{^{\prime}},\beta^{^{\prime}})=%
%TCIMACRO{\dsum }%
%BeginExpansion
{\displaystyle\sum}
%EndExpansion
(\prod_{i=0}^{d+3}\dfrac{A(j^{^{\prime}}(i),\beta^{^{\prime}}(i),j^{^{\prime}%
}(i+1),\beta^{^{\prime}}(i+1))}{(\Lambda_{N}-\lambda_{j^{^{\prime}}%
(i),\beta^{^{\prime}}(i)})})b(N,j^{^{\prime}}(d+4),\beta^{^{\prime}}(d+4)),
\end{equation}
where sum is taken under conditions $(j_{1},\beta_{1})\in Q(\rho^{\alpha
},9c),(j_{2},\beta_{2})\in Q(\rho^{\alpha},90c),...,$ $(j_{d+4},\beta
_{d+4})\in Q(\rho^{\alpha},(10^{d+3})9c).$ Now using (34), (82), and (84), we
obtain the proof of (55). It follows from (83) and (55) that the series in
(78) can be term by term differentiated and (79) holds. Substituting (79) into
(77) and using the Parseval equality by direct calculation we obtain the proof
of the lemma.

{\Large The proof of Lemma 3}

By Lemma 2 we have
\begin{equation}
\mid\beta+\tau\mid\frac{\partial\Lambda_{N}(t)}{\partial h}=\sum_{j^{^{\prime
}}\in\mathbb{Z},\beta^{^{\prime}}\in\Gamma_{\delta}}(\beta+\tau,\beta
^{^{\prime}}+\tau)\mid b(N,j^{^{\prime}},\beta^{^{\prime}})\mid^{2}=\sum
_{i=1}^{7}C_{i},
\end{equation}
where
\begin{equation}
C_{i}=\sum_{\beta^{^{\prime}}\in A_{i}}\sum_{j^{^{\prime}}\in\mathbb{Z}}%
(\beta+\tau,\beta^{^{\prime}}+\tau)\mid b(N,j^{^{\prime}},\beta^{^{\prime}%
})\mid^{2}%
\end{equation}
and $A_{i}$ is defined as follows

$A_{1}=\{\beta^{^{\prime}}\in\Gamma_{\delta}:\beta^{^{\prime}}+\tau\notin
R_{\delta}(4\rho)\},$

$A_{2}=\{\beta^{^{\prime}}\in\Gamma_{\delta}:\beta^{^{\prime}}+\tau\in
R_{\delta}(4\rho)\backslash R_{\delta}(H+\frac19\rho^{a-1})\},$

$A_{3}=\{\beta^{^{\prime}}\in\Gamma_{\delta}:\beta^{^{\prime}}+\tau
\in(R_{\delta}(H+\frac19\rho^{a-1})\backslash R_{\delta}(H+\rho^{\alpha_{d}%
-1})),\mid\beta-\beta^{^{\prime}}\mid\geq\rho^{a-2\alpha}\},$

$A_{4}=\{\beta^{^{\prime}}\in\Gamma_{\delta}:\beta^{^{\prime}}+\tau
\in(R_{\delta}(H+\frac19\rho^{a-1})\backslash R_{\delta}(H+\rho^{\alpha_{d}%
-1})),\mid\beta-\beta^{^{\prime}}\mid<\rho^{a-2\alpha}\},$

$A_{5}=\{\beta^{^{\prime}}\in\Gamma_{\delta}:\beta^{^{\prime}}+\tau
\in(R_{\delta}(H+\rho^{\alpha_{d}-1})\backslash R_{\delta}(H-\rho^{2\alpha
_{d}-1})),\mid\beta-\beta^{^{\prime}}\mid\geq\rho^{\alpha_{d}}\},$

$A_{6}=\{\beta^{^{\prime}}\in\Gamma_{\delta}:\beta^{^{\prime}}+\tau
\in(R_{\delta}(H+\rho^{\alpha_{d}-1})\backslash R_{\delta}(H-\rho^{2\alpha
_{d}-1})),\mid\beta-\beta^{^{\prime}}\mid<\rho^{\alpha_{d}}\}$

$A_{7}=\{\beta^{^{\prime}}\in\Gamma_{\delta}:\beta^{^{\prime}}+\tau\in
R_{\delta}(H-\rho^{2\alpha_{d}-1})\},$ where $H=\mid\beta+\tau\mid$ ,
$\beta\in S_{2}(\rho),$ and hence by definition of $S_{2}(\rho)$ ( see (23))
$H$ satisfies the inequalities%
\begin{equation}
\frac{1}{2}\rho<H<\frac{3}{2}\rho.
\end{equation}

First we prove that
\begin{equation}
C_{i}=O(\rho^{2-2a}),\text{ }\forall i=1,2,4,6.
\end{equation}

It follows from (55) that (90) holds for $i=1.$ To prove (90) for $i=2$ we use
(80) and show that
\begin{equation}
\lambda_{j^{^{\prime}},\beta^{^{\prime}}}-\Lambda_{N}(t)>c_{6}\rho^{a}.
\end{equation}
First let us prove (91). Since (53) holds, it follows from (8), (39) and the
relations $j\in S_{1}(\rho)$ ( see (23) and definition of $\ S_{1}(\rho)$)
that
\begin{equation}
\Lambda_{N}=H^{2}+O(\rho^{2\alpha_{1}}).
\end{equation}
If $\beta^{^{\prime}}\in A_{2}$ then using (89), \ definition of
$\lambda_{j^{^{\prime}},\beta^{^{\prime}}},$, and (39) we have
\begin{equation}
\lambda_{j^{^{\prime}},\beta^{^{\prime}}}>H^{2}+c_{7}\rho^{a}.
\end{equation}
This, (92), and the inequality $a>2\alpha_{1}$ imply (91). Now using (91),
(80), the inequalities $\mid\beta+\tau\mid<$ $\frac{3}{2}\rho$ ( see (89)),
$\mid\beta^{^{\prime}}+\tau\mid<$ $4\rho$ and Bessel inequality we obtain the
prove of (90) for $i=2.$

To prove (90) for $i=4$ we use the inequality $C_{4}<c_{8}\rho^{2}%
(C_{4,1}+C_{4,2}),$ where
\begin{align*}
C_{4,1}  &  =\sum_{\beta^{^{\prime}}\in A_{4}}(\sum_{j^{^{\prime}}:\mid
j^{^{\prime}}\delta\mid\geq\frac{1}{30}\rho^{\frac{a}{2}}}\mid b(N,j^{^{\prime
}},\beta^{^{\prime}})\mid^{2}),\\
\text{ }C_{4,2}  &  =\sum_{\beta^{^{\prime}}\in A_{4}}(\sum_{j^{^{\prime}%
}:\mid j^{^{\prime}}\delta\mid<\frac{1}{30}\rho^{\frac{a}{2}}}\mid
b(N,j^{^{\prime}},\beta^{^{\prime}})\mid^{2})
\end{align*}
and prove that
\begin{equation}
C_{4,i}=O(\rho^{-2a}),\forall i=1,2.
\end{equation}
It is clear that if $\beta^{^{\prime}}\in A_{4}$ and $\mid j^{^{\prime}}%
\delta\mid\geq\frac{1}{30}\rho^{\frac{a}{2}}$ then (93) holds. Therefore
repeating the prove of (90) for $i=2$ we get the proof of (94) for $i=1.$

Now we prove (94) for $i=2.$ It follows from (80) that
\begin{equation}
C_{4,2}=\sum_{\beta^{^{\prime}}\in A_{4}}(\sum_{j^{^{\prime}}:\mid
j^{^{\prime}}\delta\mid<\frac{1}{30}\rho^{\frac{a}{2}}}\frac{\mid(\Psi
_{N}(x),(q(x)-Q(s))\Phi_{j^{^{\prime}},\beta^{^{\prime}}}(x))\mid^{2}}%
{\mid\Lambda_{N}(t)-\lambda_{j^{^{\prime}},\beta^{^{\prime}}}\mid^{2}}.
\end{equation}
Since $\alpha_{d}>2\alpha_{1}$ it follows from (92) that the inequality
$\lambda_{j^{^{\prime}},\beta^{^{\prime}}}-\Lambda_{N}(t)>c_{9}\rho
^{\alpha_{d}}$ holds for $\beta^{^{\prime}}\in A_{4},$ $\mid j^{^{\prime}%
}\delta\mid<\rho^{\frac{a}{2}}$. Therefore using (39) we obtain
\begin{equation}
\sum_{j^{^{\prime}}:\mid j^{^{\prime}}\delta\mid<\frac{1}{30}\rho^{\frac{a}%
{2}}}\frac{1}{\mid\Lambda_{N}(t)-\lambda_{j^{^{\prime}},\beta^{^{\prime}}}%
\mid^{2}}<c_{10},\text{ }\forall\beta^{^{\prime}}\in A_{4},
\end{equation}
where $c_{10}$ does not depend on $\beta^{^{\prime}}.$ Using this \ in (95)
and denoting%
\[
\mid(\Psi_{N},(q(x)-Q(s))\Phi_{n(\beta^{^{\prime}}),\beta^{^{\prime}}}%
(x))\mid=\max_{j^{^{\prime}}:\mid j^{^{\prime}}\delta\mid<\frac{1}{30}%
\rho^{\frac{a}{2}}}\mid(\Psi_{N},(q(x)-Q(s))\Phi_{j^{^{\prime}},\beta
^{^{\prime}}}(x))\mid
\]
(if $\max$ is gotten for several index $n(\beta^{^{\prime}}),$ then we take
one of them), we get
\[
C_{4,2}<c_{11}\sum_{\beta^{^{\prime}}\in A_{4}}\mid(\Psi_{N}%
(x),(q(x)-Q(s))\Phi_{n(\beta^{^{\prime}}),\beta^{^{\prime}}}(x))\mid^{2}.
\]
Now using (33), (34) and then (80) we obtain%
\begin{equation}
C_{4,2}<c_{12}\rho^{-p\alpha}+c_{12}\sum_{\beta^{^{\prime}}\in A_{4}}\mid
b(N,n(\beta^{^{\prime}})+j_{1}(\beta^{^{\prime}}),\beta^{^{\prime}}+\beta
_{1}(\beta^{^{\prime}}))\mid^{2}%
\end{equation}%
\[
=c_{12}\rho^{-p\alpha}+c_{12}\sum_{\beta^{^{\prime}}\in A_{4}}\frac{\mid
(\Psi_{N},(q(x)-Q(s))\Phi_{n(\beta^{^{\prime}})+j_{1}(\beta^{^{\prime}}%
),\beta^{^{\prime}}+\beta_{1}(\beta^{^{\prime}})}(x))\mid^{2}}{\mid\Lambda
_{N}-\lambda_{n(\beta^{^{\prime}})+j_{1}(\beta^{^{\prime}}),\beta^{^{\prime}%
}+\beta_{1}(\beta^{^{\prime}})}\mid^{2}},
\]
where%
\[
\mid b(N,n(\beta^{^{\prime}})+j_{1}(\beta^{^{\prime}}),\beta^{^{\prime}}%
+\beta_{1}(\beta^{^{\prime}}))\mid=\max_{(j_{1},\beta_{1})\in Q(\rho^{\alpha
},9\frac{1}{30}\rho^{\frac{a}{2}})}\mid b(N,n(\beta^{^{\prime}})+j_{1}%
,\beta^{^{\prime}}+\beta_{1})\mid.
\]
To estimate $C_{4,2}$ let us prove that
\begin{equation}
\mid\Lambda_{N}-\lambda_{n(\beta^{^{\prime}})+j_{1}(\beta^{^{\prime}}%
),\beta^{^{\prime}}+\beta_{1}(\beta^{^{\prime}})}\mid>\frac{1}{8}\rho^{a}.
\end{equation}
The inclusion $(j_{1},\beta_{1})\in Q(\rho^{\alpha},9\frac{1}{30}\rho
^{\frac{a}{2}})$ and the condition $\mid j^{^{\prime}}\delta\mid<\frac{1}%
{30}\rho^{\frac{a}{2}}$ imply that $\mid n(\beta^{^{\prime}})\delta
+j_{1}(\beta^{^{\prime}})\delta\mid<\frac{1}{3}\rho^{\frac{a}{2}}$ and by
(39)
\[
\mid\mu_{n(\beta^{^{\prime}})+j_{1}(\beta^{^{\prime}})}\mid<\frac{1}{8}%
\rho^{a}.
\]
Therefore, by (92), to prove (98) it is enough to show that
\begin{equation}
\mid H^{2}-\mid\beta^{^{\prime}}+\beta_{1}+\tau\mid^{2}\mid>\frac{3}{8}%
\rho^{a},\forall\beta^{^{\prime}}\in A_{4},\beta_{1}\in\Gamma_{\delta}%
(p\rho^{\alpha}).
\end{equation}
Since $\mid\mid\beta^{^{\prime}}+\tau\mid^{2}-H^{2}\mid<\frac{1}{2}\rho^{a}$
(see definition of $A_{4}$ and use (89)) we need to prove that
\begin{equation}
\mid\mid\beta^{^{\prime}}+\tau\mid^{2}-\mid\beta^{^{\prime}}+\beta_{1}%
+\tau\mid^{2}\mid>\frac{7}{8}\rho^{a},\forall\beta^{^{\prime}}\in A_{4}%
,\beta_{1}\in\Gamma_{\delta}(p\rho^{\alpha}).
\end{equation}
Using $\mid\beta-\beta^{^{\prime}}\mid<\rho^{a-2\alpha}$ ( see definition of
$A_{4})$ by calculations we get
\begin{align*}
&  \mid\beta^{^{\prime}}+\tau\mid^{2}-\mid\beta^{^{\prime}}+\beta_{1}+\tau
\mid^{2}=-2(\beta^{^{\prime}}+\tau,\beta_{1})-\mid\beta_{1}\mid^{2}=\\
-2(\beta+\tau,\beta_{1})-  &  \mid\beta_{1}\mid^{2}-2(\beta^{^{\prime}}%
-\beta,\beta_{1})=-(\mid\beta+\beta_{1}+\tau\mid^{2}-\mid\beta+\tau\mid
^{2})+o(\rho^{a}).
\end{align*}
This and (31) imply that (100) and hence (98) holds. Now to estimate the
right-hand side of \ (97) we prove that if $\beta^{^{\prime}}\in A_{4},$
$\beta^{^{\prime\prime}}\in A_{4}$ and $\beta^{^{\prime}}\neq\beta
^{^{\prime\prime}}$ then
\begin{equation}
\beta^{^{\prime}}+\beta_{1}(\beta^{^{\prime}})\neq\beta^{^{\prime\prime}%
}+\beta_{1}(\beta^{^{\prime\prime}}).
\end{equation}
Assume the contrary that they are equal. Then we have $\beta^{^{\prime\prime}%
}=\beta^{^{\prime}}+b,$ where $b\in\Gamma_{\delta}(2\rho^{\alpha}),$ since
$\beta_{1}(\beta^{^{\prime}})\in\Gamma_{\delta}(\rho^{\alpha}),$ $\beta
_{1}(\beta^{^{\prime\prime}})\in\Gamma_{\delta}(\rho^{\alpha}).$ It easily
follows from the inclusions $\beta^{^{\prime}}\in A_{4},$ $\beta^{^{\prime}%
}+b\in A_{4}$ \ that%
\[
\ \mid\mid\beta^{^{\prime}}+\tau\mid^{2}-\mid\beta^{^{^{\prime}}}+\tau
+b\mid^{2}\mid<\frac{1}{2}\rho^{a}%
\]
which contradicts (100)). Thus (101) is proved. Therefore using (98) and
Bessel inequality from (97) we obtain the proof of (94) for $i=2.$ Hence (90)
is proved for $i=4.$

Now we prove (90) for $i=6.$ First we note that $A_{6}=\{\beta\}.$ Indeed if
$\beta^{^{\prime}}\neq\beta$ and $\beta^{^{\prime}}\in A_{6}$ then we have
$\beta^{^{\prime}}=\beta+b,$ where $b\in\Gamma_{\delta}(\rho^{\alpha_{d}}),$
and from the relations $\beta\notin V_{b}^{\delta}(\rho^{\frac{1}{2}})$ ( see
(23) and the definition of $S_{2}$), $\mid\beta+\tau\mid=H,$ we obtain that
$\mid\mid\beta^{^{\prime}}+\tau\mid^{2}-H^{2}\mid>\frac{1}{2}\rho^{\frac{1}%
{2}}$ which contradicts $\beta^{^{\prime}}+\tau\in R_{\delta}(H+\rho
^{\alpha_{d}-1})).$ Hence
\[
C_{6}=\sum_{j^{^{\prime}}\in\mathbb{Z}}(\beta+\tau,\beta+\tau)\mid
b(N,j^{^{\prime}},\beta)\mid^{2}=H^{2}\sum_{j^{^{\prime}}\in\mathbb{Z}}\mid
b(N,j^{^{\prime}},\beta)\mid^{2}=H^{2}\sum_{i=1}^{3}C_{6,i},
\]
where $C_{6,1}=\mid b(N,j,\beta)\mid^{2},$
\[
C_{6,2}=\sum_{\mid j^{^{\prime}}\delta\mid\geq\frac{1}{30}\rho^{\frac{a}{2}}%
}\mid b(N,j^{^{\prime}},\beta)\mid^{2},\text{ }C_{6,3}=\sum_{\mid j^{^{\prime
}}\delta\mid<\frac{1}{30}\rho^{\frac{a}{2}},j^{^{\prime}}\neq j}\mid
b(N,j^{^{\prime}},\beta)\mid^{2}.
\]
To prove (90) for $i=6$ we show that
\begin{equation}
C_{6,i}=O(\rho^{-2a}),\text{ }\forall i=1,2,3.
\end{equation}
By (57) this equality holds for $i=1.$ For $\mid j^{^{\prime}}\delta\mid
\geq\frac{1}{30}\rho^{\frac{a}{2}}$ the inequality (91) holds. Therefore
repeating the proof of (90) for $i=2$ we get the proof of (102) for $i=2.$
Arguing as in the proof of (94) for $i=2$ we obtain the proof of (102) for
$i=3.$ Thus (90) is proved for $i=6$.

Now we prove that
\begin{equation}
C_{i}\leq\sum_{\beta^{^{\prime}}\in A_{i}}(\sum_{j^{^{\prime}}\in Z}\mid
b(N,j^{^{\prime}},\beta^{^{\prime}})\mid^{2}(H^{2}-\frac{1}{3}\rho
^{2\alpha_{d}}))
\end{equation}
for $i=3,5,7.$ Consider the triangle generated by vectors $\beta+\tau,$
$\beta^{^{\prime}}+\tau,$ $\beta-\beta^{^{\prime}}.$ For $\beta^{^{\prime}}\in
A_{3}$ we have
\[
H+\rho^{\alpha_{d}-1}\leq\mid\beta^{^{\prime}}+\tau\mid\leq H+\frac{1}{9}%
\rho^{a-1},\mid\beta-\beta^{^{\prime}}\mid\geq\rho^{a-2\alpha}.
\]
Let $\theta$ be the angle between the vectors $\beta+\tau,$ and $\beta
^{^{\prime}}+\tau$. If $\mid\theta\mid\leq\frac{\pi}{2}$ then using the cosine
theorem we get%
\begin{align*}
&  \mid(\beta+\tau,\beta^{^{\prime}}+\tau)\mid=\frac{1}{2}(\mid\beta+\tau
\mid^{2}+\mid\beta^{^{\prime}}+\tau\mid^{2}-\mid\beta-\beta^{^{\prime}}%
\mid^{2})\\
&  \leq H^{2}-\frac{1}{3}\rho^{2a-4\alpha}<H^{2}-\frac{1}{3}\rho^{2\alpha_{d}%
},
\end{align*}
since $a-2\alpha>\alpha_{d}.$ Using this and taking into account that
$(\beta+\tau,\beta^{^{\prime}}+\tau)<0$ for $\frac{\pi}{2}<\mid\theta\mid
\leq\pi$ we get the proof of (103) for $i=3.$ Similarly if $\beta^{^{\prime}%
}\in A_{5},$ $\mid\theta\mid\leq\frac{\pi}{2}$ then%
\[
\mid(\beta+\tau,\beta^{^{\prime}}+\tau)\mid\leq H^{2}-\frac{1}{3}\rho
^{2\alpha_{d}}%
\]
and hence (103) holds for $i=5$. If $\beta^{^{\prime}}\in A_{7}$ then
$\mid\beta^{^{\prime}}+\tau\mid\leq H-\rho^{2\alpha_{d}-1}$ and by (89) we
have
\[
\mid(\beta+\tau,\beta^{^{\prime}}+\tau)\mid\leq H^{2}-\frac{1}{3}\rho
^{2\alpha_{d}},
\]
that is, (103) holds for $i=7$ too. Now (103) and Bessel inequality imply
that
\[
C_{3}+C_{5}+C_{7}\leq H^{2}-\frac{1}{3}\rho^{2\alpha_{d}}=\mid\beta+\tau
\mid^{2}-\frac{1}{3}\rho^{2\alpha_{d}}.
\]
This, (90) and (54) give the proof of Lemma 3, since $2-2a<2\alpha_{d}$ ( see
the definition of $a$ in (9)).

{\Large The proof of Lemma 4}

Let $n_{1}$ be a positive integer satisfying the inequality

$\mid(n_{1}+v)\delta\mid^{2}\leq4\rho^{1+\alpha_{d}}<\mid(n_{1}+1+v)\delta
\mid^{2}.$ Introduce the following sets%

\[
D_{b^{^{\prime}},j}(\rho,v,4)=\{x\in H_{\delta}:\mid2(x,b^{^{\prime}})+\mid
b^{^{\prime}}\mid^{2}+\mid(j+v)\delta\mid^{2}\mid<4d_{\delta}\rho^{\alpha_{d}%
}\}
\]%
\begin{equation}
D(\rho,v,4)=%
%TCIMACRO{\dbigcup \limits_{j=-n_{1}-3}^{n_{1}}}%
%BeginExpansion
{\displaystyle\bigcup\limits_{j=-n_{1}-3}^{n_{1}}}
%EndExpansion
(\bigcup_{b^{^{\prime}}\in\Gamma_{\delta}(\rho^{\alpha_{d}})}D_{b^{^{\prime}%
},j}(\rho,v,4),
\end{equation}

\begin{equation}
S_{2}^{^{\prime}}(\rho,b,v)=((V_{b}^{\delta}(4\rho^{a})\backslash
V_{b}^{\delta}(\rho^{a}))\backslash(D(\rho,v,4)\cup D_{1}(\rho^{\frac{1}{2}%
})\cup D_{2}(\rho^{a+2\alpha})))\cap D_{3},
\end{equation}
where
\[
D_{1}(\rho^{\frac{1}{2}})=\bigcup_{b^{^{\prime}}\in\Gamma_{\delta}%
(\rho^{\alpha_{d}})}V_{b^{^{\prime}}}^{\delta}(\rho^{\frac{1}{2}}),\text{
}D_{2}(\rho^{a+2\alpha})=\bigcup_{b^{^{\prime}}\in\Gamma_{\delta}%
(p\rho^{\alpha})\backslash b\mathbb{R}}V_{b^{^{\prime}}}^{\delta}%
(\rho^{a+2\alpha}),
\]

$D_{3}=(R(\frac{3}{2}\rho-d_{\delta}-1)\backslash R(\frac{1}{2}\rho+d_{\delta
}+1)).$

Now we prove that the set $S_{2}^{^{\prime}}(\rho,b,v)$ contains an element
$\beta\in\Gamma_{\delta}$ satisfying all assertions of Lemma 4. First let us
prove that $S_{2}^{^{\prime}}(\rho,b,v)\cap\Gamma_{\delta}$ is nonempty subset
of $S_{2}(\rho),$ that is,
\begin{equation}
S_{2}^{^{\prime}}(\rho,b,v)\cap\Gamma_{\delta}\subset S_{2}(\rho),\text{
}S_{2}^{^{\prime}}(\rho,b,v)\cap\Gamma_{\delta}\neq\emptyset
\end{equation}
It follows from the definitions of $S_{2}^{^{\prime}}(\rho,b,v)$ and
$S_{2}(\rho)$ ( see (23)) that the first relation of (106) holds. To prove the
second relation we consider the set%
\[
D^{^{\prime}}(\rho)=(V_{b}^{\delta}(3\rho^{a})\backslash V_{b}^{\delta}%
(2\rho^{a}))\backslash(D(\rho,v,6)\cup D_{1}(2\rho^{\frac{1}{2}})\cup
D_{2}(2\rho^{a+2\alpha})))\cap D_{4},
\]
where $D_{4}=R(\frac{3}{2}\rho-1)\backslash R(\frac{1}{2}\rho+1).$ If
$\beta+\tau\in D^{^{\prime}}(\rho),$ where $\beta\in\Gamma_{\delta},\tau\in
F_{\delta},$ then one can easily verify that $\beta\in S_{2}^{^{\prime}}%
(\rho,b,v).$ Therefore $\{$ $\beta+F_{\delta}:$ $\beta\in S_{2}^{^{\prime}%
}(\rho,b,v)\cap\Gamma_{\delta}\}$ is a cover of $D^{^{\prime}}(\rho).$ Hence
\begin{equation}
\mid S_{2}^{^{\prime}}(\rho,b,v)\cap\Gamma_{\delta}\mid\geq(\mu(F_{\delta
}))^{-1}\mu(D^{^{\prime}}(\rho)),
\end{equation}
where $\mid S_{2}^{^{\prime}}(\rho,b,v)\cap\Gamma_{\delta}\mid$ is the number
of elements of $S_{2}^{^{\prime}}(\rho,b,v)\cap\Gamma_{\delta}.$ Thus to prove
the second relation of (106) we need to estimate $\mu(D^{^{\prime}}(\rho)).$
It is not hard to verify that ( see Remark 2.1 of [4])%
\begin{equation}
\mu((V_{b}^{\delta}(3\rho^{a})\backslash V_{b}^{\delta}(2\rho^{a}))\cap
D_{4})>c_{13}\rho^{d-2+a}.
\end{equation}
Now we estimate $\mu((V_{b}^{\delta}(3\rho^{a})\backslash V_{b}^{\delta}%
(2\rho^{a}))\cap D_{1}(2\rho^{\frac{1}{2}})\cap D_{4}).$ If $b^{^{\prime}}%
\in(b\mathbb{R)\cap}\Gamma_{\delta}(\rho^{\alpha_{d}})$, then one can easily
verify that $V_{b^{^{\prime}}}^{\delta}(2\rho^{\frac{1}{2}})\cap D_{4}\subset
V_{b}^{\delta}(2\rho^{a})\cap D_{4}.$ Therefore we need to estimate the
measure of $V_{b}^{\delta}(3\rho^{a})\cap V_{b^{^{\prime}}}^{\delta}%
(2\rho^{\frac{1}{2}})\cap D_{4}$ for $b^{^{\prime}}\in\Gamma_{\delta}%
(\rho^{\alpha_{d}})\backslash b\mathbb{R}$. For this we turn the coordinate
axes so that the direction of $(1,0,0,...,0)$ coincides with the direction
$b^{^{\prime}},$ that is $b^{^{\prime}}=(\mid b^{^{\prime}}\mid,0,0,...,0)$
and the plane generated by $b,b^{^{\prime}}$ coincides with the plane
$(x_{1},x_{2},0,...0),$ that is, $b=(b_{1},b_{2},0,...,0).$Then the condition
$x\in V_{b}^{\delta}(3\rho^{a})\cap V_{b^{^{\prime}}}^{\delta}(2\rho^{\frac
{1}{2}})\cap D_{4}$ imply that
\begin{align}
x_{1}  &  \mid b^{^{\prime}}\mid=O(\rho^{\frac{1}{2}}),\nonumber\\
x_{1}b_{1}+x_{2}b_{2}  &  =O(\rho^{a}).\\
x_{1}^{2}+x_{2}^{2}+...+x_{d-1}^{2}  &  =O(\rho^{2}).\nonumber
\end{align}
First equality of (109) shows that $x_{1}=O(\rho^{\frac{1}{2}}).$ Since
$b^{^{\prime}}$ and $b$ are linearly independent vectors of $\Gamma_{\delta}$
we have $\mid b^{^{\prime}}\mid\mid b_{2}\mid\geq\mu(F_{\delta}),$ where $\mid
b^{^{\prime}}\mid<\rho^{\alpha_{d}}.$ Therefore $\mid b_{2}\mid\geq
\mu(F_{\delta})\rho^{-\alpha_{d}}$ and the second equality of (109) implies
that $x_{2}=O(\rho^{a+\alpha_{d}}).$ Now using the third equality of (109) we
obtain that $V_{b}^{\delta}(3\rho^{a})\cap V_{b^{^{\prime}}}^{\delta}%
(2\rho^{\frac{1}{2}})\cap D_{4}$ is subset of $[-c_{14}\rho^{\frac{1}{2}%
},c_{14}\rho^{\frac{1}{2}}]\times\lbrack-c_{14}\rho^{a+\alpha_{d}},c_{14}%
\rho^{a+\alpha_{d}}]\times([-c_{14}\rho,c_{14}\rho])^{d-3}$ which has the
measure $O(\rho^{d-3+\frac{1}{2}+a+\alpha_{d}}).$ This with $\mid
\Gamma_{\delta}(\rho^{\alpha_{d}})\mid=O(\rho^{(d-1)\alpha_{d}})$ give%
\begin{equation}
\mu((V_{b}^{\delta}(3\rho^{a})\cap D_{1}(2\rho^{\frac{1}{2}})\cap
D_{4})=O(\rho^{d-3+\frac{1}{2}+a+d\alpha_{d}})=o(\rho^{d-2+a}),
\end{equation}
since $d\alpha_{d}<\frac{1}{2}$ ( see the definition of $\alpha_{d}$ in (7)).
In the same way we get%
\begin{equation}
\mu((V_{b}^{\delta}(3\rho^{a})\cap D_{2}(2\rho^{a+2\alpha})\cap D_{4}%
)=O(\rho^{d-3+2a+(d+4)\alpha})=o(\rho^{d-2+a}),
\end{equation}
since $a+(d+4)\alpha<1$ ( see (7) and (9)). To estimate $\mu(D_{b^{^{\prime}%
},j}(\rho,v,6))$ we turn the coordinate axes so that the direction of
$(1,0,0,...,0)$ coincides with the direction $b^{^{\prime}}.$ Then the
condition $x\in D_{b^{^{\prime}},j}(\rho,v,6)\cap D_{4}$ imply that
\[
2x_{1}\mid b^{^{\prime}}\mid+\mid b^{^{\prime}}\mid^{2}+\mid(j+v)\delta
\mid^{2}\mid=O(\rho^{\alpha_{d}}),
\]%
\[
x_{1}^{2}+x_{2}^{2}+...+x_{d-1}^{2}=O(\rho^{2}).
\]
These equalities shows that $x_{1}$ belongs to the interval of length
$O(\rho^{\alpha_{d}})$ and%
\[
\mu(D_{b^{^{\prime}},j}(\rho,v,6)\cap D_{4})=O(\rho^{d-2+\alpha_{d}}).
\]
Now using (104) and taking into account that $n_{1}=O(\rho^{\frac{1}%
{2}(1+\alpha_{d})})$,

$\mid\Gamma_{\delta}(\rho^{\alpha_{d}})\mid=O(\rho^{(d-1)\alpha_{d}})$ we
obtain
\[
\mu(D(\rho,v,4)\cap D_{4}=O(\rho^{d-2+\frac{1}{2}+(d+\frac{1}{2})\alpha_{d}%
})=o(\rho^{d-2+a}),
\]
since $a>\frac{1}{2}+(d+\frac{1}{2})\alpha_{d}$ ( see (9) and (7)). This
estimation with \ (110), (111), and (108) implies that $\mu(D^{^{\prime}}%
(\rho))>c_{15}\rho^{d-2+a}.$ Therefore (107) give the proof of the second
equality of (106).

Now take any element $\beta$ from $S_{2}^{^{\prime}}(\rho,b,v)\cap
\Gamma_{\delta}.$ It follows from the definitions of of the sets
$S_{2}^{^{\prime}}(\rho,b,v),$ $D_{b^{^{\prime}},j}(\rho,v,4),$ $A(\beta
,\rho)$ ( see (105) and (27)) that $v\notin A(\beta,\rho).$

Let us prove the inequalities in (58). By the definition of $S_{2}^{^{\prime}%
}(\rho,b,v)$ we have $\beta\in V_{b}^{\delta}(4\rho^{a})\backslash
V_{b}^{\delta}(\rho^{a})$. This means that
\[
\rho^{a}\leq\mid2(\beta,b)+\mid b\mid^{2}\mid<4\rho^{a}.
\]
This with the obvious relations $\mid b\mid=O(1),$ $\mid\tau\mid=O(1)$ imply (58).

Now we prove (59). If $\gamma\in S(\delta,b)\backslash\delta\mathbb{R}$ then%
\begin{equation}
\gamma=nb+a\delta,\text{ }n\neq0,\text{ }n\in\mathbb{Z},\text{ }a\in
\mathbb{R},\text{ }\mid(\gamma,b)\mid=\mid n\mid\mid b\mid^{2}\geq\mid
b\mid^{2},
\end{equation}
since each $\gamma\in\Gamma$ has decomposition $\gamma=b^{^{\prime}}+a\delta,$
where $b^{^{\prime}}\in\Gamma_{\delta},$ and $b$ is a maximal element of
$\Gamma_{\delta}$ (see (3.2) of [4] and the definition of $S(\delta,b)$ in
(13)). This with the relations $(\beta+\tau,\delta)=0$ give $(\beta
+\tau,\gamma)=n(\beta+\tau,b)$. Therefore the first inequality of (58) implies
(59). \ 

Let us prove (60). If $\gamma\not \in S(\delta,b),\mid\gamma\mid
<|\rho|^{\alpha}$ then $\gamma=b^{^{\prime}}+a\delta,$ where $a\in\mathbb{R},$
$b^{^{\prime}}\in\Gamma_{\delta}(\rho^{\alpha})\backslash b\mathbb{R}$, and
$(\beta+\tau,\gamma)=(\beta+\tau,b^{^{\prime}}).$ Therefore using $\mid
b^{^{\prime}}\mid=O(\rho^{\alpha}),$ $\mid\tau\mid=O(1)$ and arguing as in the
proof of (58) we see that the relation

$\beta\notin V_{b^{^{\prime}}}^{\delta}(\rho^{a+2\alpha}),$ ( see definition
of $S_{2}^{^{\prime}}(\rho,b,v)$) implies (60).

The inequality (61) follows from the definition of $f_{\delta,\beta+\tau}(x)$,
(59), (60), and from the obvious relation
\[
\sum_{\gamma\in\Gamma}\mid\gamma\mid\mid q_{\gamma}\mid<c_{16},\forall q(x)\in
W_{2}^{s}(F,M).
\]
The last inequality with (112) imply the convergence of the series (13).

\section{APPENDICES}

\ \ \ \ {\large APPENDIX A. THE PROOF OF (76).}\textbf{ }

Here we estimate the complex conjugate $\overline{C_{1}(j^{^{\prime}}%
,\lambda_{j,\beta})}$ of $C_{1}(j^{^{\prime}},\lambda_{j,\beta}),$ namely
prove that (see (74))%
\begin{equation}
\sum_{(j_{1},\beta_{1})\in Q(\rho^{\alpha},9r)}\dfrac{\overline{A(j^{^{\prime
}},\beta,j^{^{\prime}}+j_{1,}\beta+\beta_{1}})\overline{A(j^{^{\prime}}%
+j_{1,}\beta+\beta_{1},j,\beta})}{\lambda_{j,\beta}-\lambda_{j^{^{\prime}%
}+j_{1},\beta+\beta_{1}}}=O(\rho^{-2a}r^{2}), \tag{A1}%
\end{equation}
where $Q(\rho^{\alpha},9r)=\{(j_{1},\beta_{1}):\mid j_{1}\delta\mid<9r,$
$0<\mid\beta_{1}\mid<\rho^{\alpha}\}$, $j\in S_{1}(\rho),$ $\mid j^{^{\prime}%
}\delta\mid<r$, $r=O(\rho^{\frac{1}{2}\alpha_{2}}).$ The conditions on indices
$j^{^{\prime}}$, $j_{1},$ $j$ and (39) imply that

$\mu_{j^{^{\prime}}+j_{1}}=O(r^{2}),$ $\mu_{j}=O(r^{2}).$ These with
$\beta\notin V_{\beta_{1}}^{\delta}(\rho^{a})))$, where $\beta_{1}\in
\Gamma_{\delta}(p\rho^{\alpha}),$ ( see (29)) give
\begin{equation}
\lambda_{j,\beta}-\lambda_{j^{^{\prime}}+j_{1},\beta+\beta_{1}}=-2(\beta
,\beta_{1})+O(r^{2}),\text{ }\mid(\beta,\beta_{1})\mid>\frac{1}{3}\rho^{a}.
\tag{A2}%
\end{equation}
Using \ this, (34) and (A1) we get
\begin{equation}
\overline{C_{1}(j^{^{\prime}},\lambda_{j,\beta})}=\sum_{\beta_{1}}%
\dfrac{C^{^{\prime}}}{-2(\beta,\beta_{1})}+O(\rho^{-2a}r^{2}), \tag{A3}%
\end{equation}
where $C^{^{\prime}}=\sum_{j_{1}}\overline{A(j^{^{\prime}},\beta,j^{^{\prime}%
}+j_{1,}\beta+\beta_{1}})\overline{A(j^{^{\prime}}+j_{1,}\beta+\beta
_{1},j,\beta}).$ In [4] we proved that ( see (3.21), (3.7), Lemma 3.3 of [4])%

\begin{align}
\overline{A(j^{^{\prime}},\beta,j^{^{\prime}}+j_{1,}\beta+\beta_{1}})  &
=\sum\limits_{n_{1}:(n_{1},\beta_{1})\in\Gamma^{^{\prime}}(\rho^{\alpha}%
)}c(n_{1},\beta_{1})a(n_{1},\beta_{1},j^{^{\prime}},\beta,j^{^{\prime}}%
+j_{1,}\beta+\beta_{1}),\tag{A4}\\
\overline{A(j^{^{\prime}}+j_{1,}\beta+\beta_{1},j,\beta})  &  =\sum
\limits_{n_{2}:(n_{2},-\beta_{1})\in\Gamma^{^{\prime}}(\rho^{\alpha})}%
c(n_{2},-\beta_{1})a(n_{2},-\beta_{1},j^{^{\prime}}+j_{1,}\beta+\beta
_{1},j,\beta),\nonumber
\end{align}

$\Gamma^{^{\prime}}(\rho^{\alpha})=\{(n_{1},\beta_{1}):\beta_{1}\in
\Gamma_{\delta}\backslash0,n_{1}\in\mathbb{Z},\beta_{1}+(n_{1}-(2\pi
)^{-1}(\beta_{1},\delta^{\ast}))\delta\in\Gamma(\rho^{\alpha})\},$
\begin{equation}
c(n_{1},\beta_{1})=q_{\gamma_{1}},\gamma_{1}=\beta_{1}+(n_{1}-(2\pi
)^{-1}(\beta_{1},\delta^{\ast}))\delta\in\Gamma(\rho^{\alpha}), \tag{A5}%
\end{equation}

$a(n_{1},\beta_{1},j^{^{\prime}},\beta,j^{^{\prime}}+j_{1,}\beta+\beta
_{1})=(e^{i(n_{1}-(2\pi)^{-1}(\beta_{1},\delta^{\ast}))s}\varphi_{j^{^{\prime
}},v(\beta)}(s),\varphi_{j^{^{\prime}}+j_{1},v(\beta+\beta_{1})}(s)),$%

\[
a(n_{2},-\beta_{1},j^{^{\prime}}+j_{1,}\beta+\beta_{1},j,\beta)=(e^{i(n_{2}%
-(2\pi)^{-1}(-\beta_{1},\delta^{\ast}))s}\varphi_{j^{^{\prime}}+j_{1}%
,v(\beta+\beta_{1})},\varphi_{j,v(\beta)})
\]%
\begin{equation}
=(\varphi_{j^{^{\prime}}+j_{1},v(\beta+\beta_{1})},e^{-i(n_{2}-(2\pi
)^{-1}(-\beta_{1},\delta^{\ast}))s}\varphi_{j,v(\beta)}) \tag{A6}%
\end{equation}%
\[
=\overline{(e^{-i(n_{2}-(2\pi)^{-1}(-\beta_{1},\delta^{\ast})s}\varphi
_{j,v(\beta)},\varphi_{j^{^{\prime}}+j_{1},v(\beta+\beta_{1})}),}%
\]
where $\delta^{\ast}$ is the element of $\Omega$ satisfying $(\delta^{\ast
},\delta)=2\pi$

Now to estimate the right-hand side of (A3) we prove that
\begin{align}
&  \sum_{j_{1}}a(n_{1},\beta_{1},j^{^{\prime}},\beta,j^{^{\prime}}+j_{1,}%
\beta+\beta_{1})a(n_{2},-\beta_{1},j^{^{\prime}}+j_{1},\beta+\beta_{1}%
,j,\beta)\tag{A7}\\
&  =a(n_{1}+n_{2},0,j^{^{\prime}},\beta,j,\beta)+O(\rho^{-p\alpha}).\nonumber
\end{align}
By definition we have
\[
a(n_{1}+n_{2},0,j^{^{\prime}},\beta,j,\beta)=(e^{i(n_{1}+n_{2})s}%
\varphi_{j^{^{\prime}},v(\beta)}(s),\varphi_{j,v(\beta)}(s))=
\]

$(e^{i(n_{1}-(2\pi)^{-1}(\beta_{1},\delta^{\ast}))s}\varphi_{j^{^{\prime}%
},v(\beta)}(s),e^{-i(n_{2}-(2\pi)^{-1}(-\beta_{1},\delta^{\ast}))s}%
\varphi_{j,v(\beta)}(s)).$

This, (A6), and the following formulas
\begin{align*}
&  e^{i(n_{1}-(2\pi)^{-1}(\beta_{1},\delta^{\ast}))s}\varphi_{j^{^{\prime}%
},v(\beta)}(s)\\
&  =\sum_{\mid j_{1}\delta\mid<9r}a(n_{1},\beta_{1},j^{^{\prime}}%
,\beta,j^{^{\prime}}+j_{1,}\beta+\beta_{1})\varphi_{j^{^{\prime}}%
+j_{1},v(\beta+\beta_{1})}(s)+O(\rho^{-p\alpha}),\\
&  e^{-i(n_{2}-(2\pi)^{-1}(-\beta_{1},\delta^{\ast}))s}\varphi_{j,v(\beta
)}(s)\\
&  =\sum_{\mid j_{1}\delta\mid<9r}\overline{a(n_{2},-\beta_{1},j^{^{\prime}%
},\beta,j^{^{\prime}}+j_{1,}\beta+\beta_{1})}\varphi_{j^{^{\prime}}%
+j_{1},v(\beta+\beta_{1})}+O(\rho^{-p\alpha}),
\end{align*}%
\begin{equation}
\sum\limits_{j_{_{1}}}\mid a(n_{1},\beta_{1},j^{^{\prime}},\beta,j^{^{\prime}%
}+j_{1,}\beta+\beta_{1})\mid=O(1) \tag{A8}%
\end{equation}
( see (3.16), (3.17) of [4]) give the proof of (A7). Now from (A7), (A4), (A3)
we obtain%
\[
C^{^{\prime}}=\sum\limits_{n_{1}}(\sum\limits_{n_{2}}(c(n_{1},\beta
_{1})c(n_{2},-\beta_{1})a(n_{1}+n_{2},0,j^{^{\prime}},\beta,j,\beta
)+O(\rho^{-p\alpha}))),
\]%
\[
\overline{C_{1}(j^{^{\prime}},\lambda_{j,\beta})}=\sum_{\beta_{1}}%
(\sum\limits_{n_{1}}(\sum\limits_{n_{2}}C_{1}^{^{\prime}}(\beta_{1}%
,n_{1},n_{2}))+O(\rho^{-2a}r^{2}),
\]
where $C_{1}^{^{\prime}}(\beta_{1},n_{1},n_{2})=\dfrac{c(n_{1},\beta
_{1})c(n_{2},-\beta_{1})a(n_{1}+n_{2},0,j^{^{\prime}},\beta,j,\beta)}%
{-2(\beta,\beta_{1})}.$ One can readily verify that
\begin{equation}
C_{1}^{^{\prime}}(\beta_{1},n_{1},n_{2})+C_{1}^{^{\prime}}(-\beta_{1}%
,n_{2},n_{1})=0. \tag{A9}%
\end{equation}
Therefore $\overline{C_{1}(j^{^{\prime}},\lambda_{j,\beta})}=O(\rho^{-2a}%
r^{2}).$

\ \ \ \ {\large APPENDIX B. THE PROOF OF (47).}

Arguing as in the proof of (75) we see that%
\[
C_{2}(\Lambda_{j,\beta})=C_{2}(\lambda_{j,\beta})+O(\rho^{-3a}).
\]
Using (A4) we obtain
\begin{align*}
\overline{C_{2}(\lambda_{j,\beta})}  &  =\sum_{\beta_{1},\beta_{2}}%
(\sum\limits_{n_{1},n_{2},n_{3}}(\sum_{j_{1},j_{2}}\dfrac{c(n_{1},\beta
_{1})c(n_{2},\beta_{2})c(n_{3},-\beta_{1}-\beta_{2})}{(\lambda_{j,\beta
}-\lambda_{j(1),\beta(1)})(\lambda_{j,\beta}-\lambda_{j(2)_{,}\beta(2)}%
)}a(n_{1},\beta_{1},j,\beta,j(1)_{,}\beta(1))\times\\
&  a(n_{2},\beta_{2},j(1)_{,}\beta(1),j(2)_{,}\beta(2))a(n_{3},-\beta
_{1}-\beta_{2},j(2),\beta(2),j,\beta),
\end{align*}
where $(j_{1},\beta_{1})\in Q(\rho^{\alpha},9r_{1}),$ $(j_{2},\beta_{2})\in
Q(\rho^{\alpha},90r_{1}),$ $j\in S_{1},\beta_{1}+\beta_{2}\neq0.$ Applying
(A7) two times and using (A8), we get%

\[
\sum_{j_{1}}a(n_{1},\beta_{1},j,\beta,j(1)_{,}\beta(1))(\sum_{j_{2}}%
a(n_{2},\beta_{2},j(1)_{,}\beta(1),j(2)_{,}\beta(2))a(n_{3},-\beta_{1}%
-\beta_{2},j(2)_{,}\beta(2),j,\beta))
\]%
\[
=\sum_{j_{1}}a(n_{1},\beta_{1},j,\beta,j(1)_{,}\beta(1))(a(n_{2}+n_{3}%
,-\beta_{1},j(1)_{,}\beta(1),j,\beta)+O(\rho^{-p\alpha}))
\]%
\[
=a(n_{1}+n_{2}+n_{3},0,j,\beta,j,\beta)+O(\rho^{-p\alpha}).
\]

Using this in above expression for $C_{2}(\lambda_{j,\beta})$ and taking into
account that
\begin{align*}
\lambda_{j,\beta}-\lambda_{j(1)_{,}\beta(1)}  &  =-2(\beta,\beta_{1}%
)+O(\rho^{2\alpha_{1}}),\mid(\beta,\beta_{1})\mid>\frac{1}{3}\rho^{a},\\
\lambda_{j,\beta}-\lambda_{j(2)_{,}\beta(2)}  &  =-2(\beta,\beta_{1}+\beta
_{2})+O(\rho^{2\alpha_{1}}),\mid(\beta,\beta_{1}+\beta_{2})\mid>\frac{1}%
{3}\rho^{a},
\end{align*}
which can be proved as (A2), we have $C_{2}(\lambda_{j,\beta})=O(\rho
^{-3a+2\alpha_{1}})+$%
\[
\sum_{\beta_{1},\beta_{2}}(\sum\limits_{n_{1},n_{2},n_{3}}\frac{c(n_{1}%
,\beta_{1})c(n_{2},\beta_{2})c(n_{3},-\beta_{1}-\beta_{2})a(n_{1}+n_{2}%
+n_{3},0,j,\beta,j,\beta)}{4(\beta,\beta_{1})(\beta,\beta_{1}+\beta_{2})}.
\]

Grouping in the last sum terms with the equal multiplicands
\[
c(n_{1},\beta_{1})c(n_{2},\beta_{2})c(n_{3},-\beta_{1}-\beta_{2}),\text{
}c(n_{2},\beta_{2})c(n_{1},\beta_{1})c(n_{3},-\beta_{1}-\beta_{2}),
\]%
\[
c(n_{1},\beta_{1})c(n_{3},-\beta_{1}-\beta_{2})c(n_{2},\beta_{2}),\text{
}c(n_{2},\beta_{2})c(n_{3},-\beta_{1}-\beta_{2})c(n_{1},\beta_{1}),
\]%
\[
c(n_{3},-\beta_{1}-\beta_{2})c(n_{1},\beta_{1})c(n_{2},\beta_{2}),\text{
}c(n_{3},-\beta_{1}-\beta_{2})c(n_{2},\beta_{2})c(n_{1},\beta_{1})
\]
and using the obvious equality
\[
\frac{1}{(\beta,\beta_{1})(\beta,\beta_{1}+\beta_{2})}+\frac{1}{(\beta
,\beta_{2})(\beta,\beta_{2}+\beta_{1})}+\frac{1}{(\beta,\beta_{1}%
)(\beta,-\beta_{2})}+
\]%
\[
\frac{1}{(\beta,\beta_{2})(\beta-,\beta_{1})}+\frac{1}{(\beta,-\beta_{1}%
-\beta_{2})(\beta,-\beta_{2})}+\frac{1}{(\beta,-\beta_{1}-\beta_{2}%
)(\beta,-\beta_{1})}=0
\]

we see that this sum is zero, that is, $C_{2}(\lambda_{j,\beta})=O(\rho
^{-3a+2\alpha_{1}}).$

{\large APPENDIX C. THE PROOF OF (46).}

It follows from (75) that $C_{1}(\Lambda_{j,\beta})=C_{1}(\lambda_{j,\beta
})+O(\rho^{-3a}).$ Therefore we need to prove that
\[
\overline{C_{1}(\lambda_{j,\beta})}=\frac{1}{4}\int_{F}\left\vert
f_{\delta,\beta+\tau}(x)\right\vert ^{2}\left\vert \varphi_{j,v}^{\delta
}\right\vert ^{2}dx+O(\rho^{-3a+2\alpha_{1}}),
\]
where%

\[
\overline{C_{1}(\lambda_{j,\beta})}\equiv\sum_{\beta_{1}}(\sum_{j_{1}}%
\dfrac{\overline{A(j,\beta,j+j_{1,}\beta+\beta_{1}})\overline{A(j+j_{1,}%
\beta+\beta_{1},j,\beta})}{\lambda_{j,\beta}-\lambda_{j+j_{1},\beta+\beta_{1}%
}},
\]
$(j_{1},\beta_{1})\in Q(\rho^{\alpha},9r_{1}),$ $j\in S_{1},$ and by (A4)%
\[
\overline{C_{1}(\lambda_{j,\beta})}=\sum_{\beta_{1}}(\sum\limits_{n_{1}%
:(n_{1},\beta_{1})\in\Gamma^{^{\prime}}(\rho^{\alpha})}(\sum\limits_{n_{2}%
:(n_{2},-\beta_{1})\in\Gamma^{^{\prime}}(\rho^{\alpha})}(\sum_{j_{1}}%
\frac{c(n_{1},\beta_{1})c(n_{2},-\beta_{1})}{\lambda_{j,\beta}-\lambda
_{j+j_{1},\beta+\beta_{1}}}\times
\]%
\[
a(n_{1},\beta_{1},j,\beta,j+j_{1,}\beta+\beta_{1})a(n_{2},-\beta_{1}%
,j+j_{1},\beta+\beta_{1},j,\beta).
\]
Replacing $\lambda_{j,\beta}-\lambda_{j+j_{1},\beta+\beta_{1}}$ by
$-(2(\beta+\tau,\beta_{1})+\mid\beta_{1}\mid^{2}+\mu_{j+j_{1}}(v(\beta
+\beta_{1}))-\mu_{j}(v(\beta)))$ and using (A7) for $j^{^{\prime}}=j$ we have
\[
\overline{C_{1}(j,\lambda_{j,\beta})}=\sum_{\beta_{1}}(\sum\limits_{n_{1}%
}(\sum\limits_{n_{2}}\frac{c(n_{1},\beta_{1})c(n_{2},-\beta_{1})a(n_{1}%
+n_{2},0,j,\beta,j,\beta)}{-2(\beta+\tau,\beta_{1})}+
\]%
\[
\sum_{\beta_{1}}(\sum\limits_{n_{1}}(\sum\limits_{n_{2}}(\sum_{j_{1}}%
\frac{c(n_{1},\beta_{1})c(n_{2},-\beta_{1})a(n_{1},\beta_{1},j,\beta
,j+j_{1,}\beta+\beta_{1})}{2(\beta+\tau,\beta_{1})(2(\beta+\tau,\beta
_{1})+\mid\beta_{1}\mid^{2}+\mu_{j+j_{1}}-\mu_{j})}\times
\]%
\[
a(n_{2},-\beta_{1},j+j_{1},\beta+\beta_{1},j,\beta)(\mid\beta_{1}\mid^{2}%
+\mu_{j+j_{1}}(v(\beta+\beta_{1}))-\mu_{j}(v(\beta))).
\]

The formula (A9) shows that the first summation of the right-hand side of this
equality is zero. Thus we need to estimate the second sum. For this we use the
following relation
\[
\mu_{j+j_{1}}(v(\beta+\beta_{1}))a(n_{1},\beta_{1},j,\beta,j+j_{1},\beta
+\beta_{1})=(e^{i(n_{1}-(2\pi)^{-1}(\beta_{1},\delta^{\ast}))s}\varphi
_{j,v(\beta)},T_{v}\varphi_{j+j_{1},v(\beta+\beta_{1})})
\]%
\[
=(T_{v}(e^{i(n_{1}-(2\pi)^{-1}(\beta_{1},\delta^{\ast}))s}\varphi_{j,v(\beta
)}(s)),\varphi_{j+j_{1},v(\beta+\beta_{1})}(s)
\]%
\[
=((\mid n_{1}-(2\pi)^{-1}(\beta_{1},\delta^{\ast})\mid^{2}\mid\delta\mid
^{2}+\mu_{j}(v))(e^{i(n_{1}-(2\pi)^{-1}(\beta_{1},\delta^{\ast}))s}%
\varphi_{j,v(\beta)}),\varphi_{j+j_{1},v(\beta+\beta_{1})})
\]%
\[
-2i(n_{1}-(2\pi)^{-1}(\beta_{1},\delta^{\ast}))\mid\delta\mid^{2}%
(e^{i(n_{1}-(2\pi)^{-1}(\beta_{1},\delta^{\ast}))s}\varphi_{j,v(\beta
)}^{^{\prime}}(s)),\varphi_{j+j_{1},v(\beta+\beta_{1})}(s)).
\]
Using this, (A7), and the formula
\begin{align*}
&  \sum_{j_{1}}(e^{i(n_{1}-(2\pi)^{-1}(\beta_{1},\delta^{\ast}))s}%
\varphi_{j,v(\beta)}^{^{\prime}}(s)),\varphi_{j+j_{1},v(\beta+\beta_{1}%
)}(s))a(n_{2},-\beta_{1},j+j_{1},\beta+\beta_{1},j,\beta)\\
&  =(e^{i(n_{1}+n_{2})s}\varphi_{j,v(\beta)}^{^{\prime}}(s)),\varphi
_{j,v(\beta)}(s))+O(\rho^{-p\alpha}),
\end{align*}
which can be proved as (A7), we obtain
\begin{equation}
\sum_{j_{1}}\mu_{j+j_{1}}(v(\beta+\beta_{1}))a(n_{1},\beta_{1},j,\beta
,j+j_{1},\beta+\beta_{1})a(n_{2},-\beta_{1},j+j_{1},\beta+\beta_{1}%
,j,\beta)=\nonumber
\end{equation}%
\begin{equation}
(\mid n_{1}-(2\pi)^{-1}(\beta_{1},\delta^{\ast})\mid^{2})\mid\delta\mid
^{2}+\mu_{j}(v))a(n_{1}+n_{2},0,j,\beta,j,\beta)- \tag{C1}%
\end{equation}%
\[
2i(n_{1}-(2\pi)^{-1}(\beta_{1},\delta^{\ast}))\mid\delta\mid^{2}%
(e^{i(n_{1}+n_{2})s}\varphi_{j,v(\beta)}^{^{\prime}}(s)),\varphi_{j,v(\beta
)}(s)).
\]
Here the last multiplicand can be estimated as follows
\[
\ \ \mu_{j}(v)(\varphi_{j,v(\beta)}(s),e^{i(n_{1}+n_{2})s}\varphi_{j,v(\beta
)}(s))=(\varphi_{j,v(\beta)}(s),T_{v}(e^{i(n_{1}+n_{2})s}\varphi_{j,v(\beta
)}(s)))
\]%
\[
=(n_{1}+n_{2})^{2}\mid\delta\mid^{2}(\varphi_{j,v(\beta)}(s),e^{i(n_{1}%
+n_{2})s}\varphi_{j,v(\beta)}(s))+
\]%
\[
2i(n_{1}+n_{2})\mid\delta\mid^{2}(\varphi_{j,v(\beta)}(s),e^{i(n_{1}+n_{2}%
)s}\varphi_{j,v(\beta)}^{^{\prime}}(s))+\mu_{j}(v)(\varphi_{j,v(\beta
)},e^{i(n_{1}+n_{2})s}\varphi_{j,v(\beta)})
\]
and hence
\[
(e^{i(n_{1}+n_{2})s}\varphi_{j,v(\beta)}^{^{\prime}}(s)),\varphi_{j,v(\beta
)}(s))=\frac{n_{1}+n_{2}}{2i}(e^{i(n_{1}+n_{2})s}\varphi_{j,v(\beta
)}(s)),\varphi_{j,v(\beta)}(s)).
\]
Using this, (C1), and (A7) we get
\begin{align*}
&  \sum_{j_{1}}(a(n_{1},\beta_{1},j,\beta,j+j_{1,}\beta+\beta_{1}%
)a(n_{2},-\beta_{1},j+j_{1},\beta+\beta_{1},j,\beta))\times\\
(  &  \mid\beta_{1}\mid^{2}+\mu_{j+j_{1}}(v(\beta+\beta_{1}))-\mu_{j}%
(v(\beta)))=a(n_{1}+n_{2},0,j,\beta,j,\beta)\times\\
(  &  \mid\beta_{1}\mid^{2}+\mid n_{1}-\frac{(\beta_{1},\delta^{\ast})}{2\pi
}\mid^{2}\mid\delta\mid^{2}-(n_{1}-\frac{(\beta_{1},\delta^{\ast})}{2\pi}%
)\mid\delta\mid^{2}(n_{1}+n_{2}))\\
\  &  =(\mid\beta_{1}\mid^{2}+\mid\delta\mid^{2}(n_{1}-\frac{(\beta_{1}%
,\delta^{\ast})}{2\pi})(-n_{2}-\frac{(\beta_{1},\delta^{\ast})}{2\pi}%
))a(n_{1}+n_{2},0,j,\beta,j,\beta).
\end{align*}
Thus $\overline{C_{1}(j,\lambda_{j,\beta})}=C+O(\rho^{-3a+2\alpha_{1}}),$
where
\begin{align}
C  &  =\sum_{\beta_{1},n_{1},n_{2}}\frac{c(n_{1},\beta_{1})c(n_{2},-\beta
_{1})a(n_{1}+n_{2},0,j,\beta,j,\beta)}{4\mid(\beta+\tau,\beta_{1})\mid^{2}%
}\times\tag{C2}\\
(  &  \mid\beta_{1}\mid^{2}+(n_{1}-\frac{(\beta_{1},\delta^{\ast})}{2\pi
})(-n_{2}-\frac{(\beta_{1},\delta^{\ast})}{2\pi})\mid\delta\mid^{2}).\nonumber
\end{align}
Now we consider
\[
\int_{F}\left\vert f_{\delta,\beta+\tau}(x)\right\vert ^{2}\left\vert
\varphi_{n,v}((\delta,x))\right\vert ^{2}dx,
\]
where $f_{\delta,\beta+\tau}(x)$ is defined in (9) and by (A5)
\[
f_{\delta,\beta+\tau}(x)=\sum_{(n_{1},\beta_{1})\in\Gamma_{\delta}^{^{\prime}%
}(\rho^{\alpha})}\frac{\beta_{1}+(n_{1}-\frac{(\beta_{1},\delta^{\ast})}{2\pi
})\delta}{(\beta+\tau,\beta_{1})}c(n_{1},\beta_{1})e^{i(\beta_{1}+(n_{1}%
-\frac{(\beta_{1},\delta^{\ast})}{2\pi})\delta,x)}.
\]
Here $f_{\delta,\beta+\tau}(x)$ is a vector of $\mathbb{R}^{d}$ and
$\left\vert f_{\delta,\beta+\tau}(x)\right\vert $ is a norm of this vector.
Using $(\beta,\delta)=0$ for $\beta\in\Gamma_{\delta}$ we obtain
\begin{align*}
\left\vert f_{\delta,\beta+\tau}(x)\right\vert ^{2}  &  =\sum_{(n_{1}%
,\beta_{1}),(n_{2},\beta_{2})\in\Gamma_{\delta}^{^{\prime}}(\rho^{\alpha}%
)}\frac{(\beta_{1},\beta_{2})+(n_{1}-\frac{(\beta_{1},\delta^{\ast})}{2\pi
})(n_{2}-\frac{(\beta_{1},\delta^{\ast})}{2\pi})\mid\delta\mid^{2}}%
{(\beta+\tau,\beta_{1})(\beta+\tau,\beta_{2})}\times\\
&  c(n_{1},\beta_{1})c(-n_{2},-\beta_{2})e^{i(\beta_{1}-\beta_{2}+(n_{1}%
-n_{2}-(2\pi)^{-1}(\beta_{1}-\beta_{2},\delta^{\ast}))\delta,x)}.
\end{align*}
Since $\varphi_{j,v}((\delta,x))$ is a function of $(\delta,x)$ we have%
\[
\int_{F}e^{i(\beta_{1}-\beta_{2}+(n_{1}-n_{2}-(2\pi)^{-1}(\beta_{1}-\beta
_{2},\delta^{\ast}))\delta,x)}\left\vert \varphi_{j,v}((\delta,x))\right\vert
^{2}dx=0
\]
for $\beta_{1}\neq\beta_{2}$. Therefore
\[
\int_{F}\left\vert f_{\delta,\beta+\tau}(x)\right\vert ^{2}\left\vert
\varphi_{j,v}((\delta,x))\right\vert ^{2}dx=\sum_{\beta_{1},n_{1},n_{2}}%
\frac{c(n_{1},\beta_{1})c(-n_{2},-\beta_{1})}{\mid(\beta+\tau,\beta_{1}%
)\mid^{2}}\times
\]%
\[
(\mid\beta_{1}\mid^{2}+(n_{1}-\frac{(\beta_{1},\delta^{\ast})}{2\pi}%
)(n_{2}-\frac{(\beta_{1},\delta^{\ast})}{2\pi})\mid\delta\mid^{2}a(n_{1}%
-n_{2},0,j,\beta,j,\beta).
\]
Replacing $n_{2}$ by $-n_{2}$ we get
\[
\int_{F}\left\vert f_{\delta,\beta+\tau}(x)\right\vert ^{2}\left\vert
\varphi_{n,v}((\delta,x))\right\vert ^{2}dx=4C.
\]
( see (C2)). Thus (46) is proved.

\ \ \ \ \ {\large APPENDIX D. ASYMPTOTIC FORMULAS FOR }$T_{v}(Q).$

Let $\mu_{0}(0)\leq\mu_{1}(0)\leq\mu_{2}(0)\leq...$ be the eigenvalues of the
operator $T_{0}(Q).$ It is well-known that $\mu_{2m+1}(0)$ and $\mu_{2m+2}(0)$
both satisfy
\[
\sqrt{\mu}=\mid(m+1)\delta\mid+\frac{1}{16\pi\mid(m+1)\delta\mid^{3}}\int
_{F}\left\vert q^{\delta}(x)\right\vert ^{2}dx+O(\frac{1}{m^{4}})
\]
(see [1], page 58). This formula yields the invariant (17). Using the
asymptotic formulas for solutions of the Sturm-Liouville equation (see [1],
page 63) one can easily obtain that
\[
\varphi_{n,v}(s)=e^{i(n+v)s}(1+\frac{Q_{1}(s)}{2i(n+v)\mid\delta\mid^{2}%
}+\frac{Q(s)-Q(0)-\frac{1}{2}Q_{1}^{2}(s)}{4(n+v)^{2}\mid\delta\mid^{4}%
})+O(\frac{1}{n^{3}})),
\]
where $Q_{1}(s)=\int_{0}^{s}Q(t)dt.$ From this by direct calculations we find
$A_{0}(\zeta),$ $A_{1}(\zeta),$ $A_{2}(\zeta)$ (see (14)) and then using these
in (15) we get the invariants (16).

Now we consider the eigenfunction $\varphi_{n,v}(s)$ of $T_{v}(p)$ in case
$v\neq0,\ \frac{1}{2}$ and%

\begin{equation}
p(s)=\sum_{m=1,-1}p_{m}e^{ims}. \tag{D1}%
\end{equation}
The eigenvalues and eigenfunctions of $T_{v}(0)$ are $\ (n+v)^{2}\mid
\delta\mid^{2}$ and $\ e^{i(n+v)s}$, for $n\in\mathbb{Z}$. Since the
eigenvalues of $T_{v}(p)$ are simple for $v\neq0,\ \frac{1}{2}$ by well-known
perturbation formula

$(\varphi_{n,v}(s),\ e^{i(n+v)s})\varphi_{n,v}(s)=e^{i(n+v)s}+$%
\begin{equation}
\sum_{k=1,2,...}\frac{(-1)^{k+1}}{2i\pi}\int\limits_{C}(T_{v}(0)-\lambda
)^{-1}p(x))^{k}(T_{v}(0)-\lambda)^{-1}e^{i(n+v)s}d\lambda, \tag{D2}%
\end{equation}
where $C$ is a contour containing only the eigenvalue $(n+t)^{2}\mid\delta
\mid^{2}$. Using
\[
(T_{v}(0)-\lambda)^{-1}e^{i(n+v)s}=\frac{e^{i(n+v)s}}{(n+v)^{2}\mid\delta
\mid^{2}-\lambda}%
\]
and (D1) we see that the $k$-th ( $k=1,2,3,4$) term $F_{k}$ of the series (D2)
has the form
\[
F_{1}=\frac{1}{2i\pi}\int\limits_{C}\sum_{m=1,-1}\frac{p_{m}e^{i(n+m+v)s}%
}{((n+v)^{2}\mid\delta\mid^{2}-\lambda)((n+m+v)^{2}\mid\delta\mid^{2}%
-\lambda)}d\lambda,
\]

\[
F_{2}=\frac{-1}{2i\pi}\int\limits_{C}\sum_{m,l=1,-1}\frac{p_{m}p_{l}%
e^{i(n+m+l+v)s}}{((n+v)^{2}\mid\delta\mid^{2}-\lambda)}\times
\]
\[
\frac1{((n+m+v)^{2}\mid\delta\mid^{2}-\lambda)((n+m+l+v)^{2}\mid\delta\mid
^{2}-\lambda)}d\lambda,
\]

\[
F_{3}=\frac1{2i\pi}\int\limits_{C}\sum_{m,l,k=1,-1}\frac{p_{m}p_{l}%
p_{k}e^{i(n+m+l+k+v)s}}{((n+v)^{2}\mid\delta\mid^{2}-\lambda)((n+m+v)^{2}%
\mid\delta\mid^{2}-\lambda)}\times
\]
\[
\frac1{((n+m+l+v)^{2}\mid\delta\mid^{2}-\lambda)((n+m+l+k+v)^{2}\mid\delta
\mid^{2}-\lambda)}d\lambda,
\]

\begin{align*}
F_{4}  &  =\frac{-1}{2i\pi}\int\limits_{C}\sum_{m,l,k,r=1,-1}\frac{p_{m}%
p_{l}p_{k}p_{r}e^{i(n+m+l+k+r+v)s}}{((n+m+l+k+r+v)^{2}\mid\delta\mid
^{2}-\lambda)}\times\\
&  \frac1{((n+m+v)^{2}\mid\delta\mid^{2}-\lambda)((n+m+l+v)^{2}\mid\delta
\mid^{2}-\lambda)}\times\\
&  \frac1{((n+m+l+k+v)^{2}\mid\delta\mid^{2}-\lambda)((n+v)^{2}\mid\delta
\mid^{2}-\lambda)}d\lambda.
\end{align*}

Since the distance between $(n+v)^{2}\mid\delta\mid^{2}$ and $(n^{^{\prime}%
}+v)^{2}\mid\delta\mid^{2}$ for $n^{^{\prime}}\neq n$ is greater than
$c_{17}n$ , we can choose the contour $C$ such that
\[
\frac{1}{\mid(n^{^{\prime}}+v)^{2}\mid\delta\mid^{2}-\lambda\mid}<\frac
{c_{18}}{n},\text{ }\forall\lambda\in C,\text{ }\forall n^{^{\prime}}\neq n
\]
and the length of $C$ is less than $c_{19}$. Therefore%
\[
\sum_{k=5,6,...}F_{k}=O(\frac{1}{n^{5}}).
\]
Now we calculate the integrals in $F_{1}$, $F_{2}$, $F_{3}$, $F_{4}$ by Cauchy
integral formula and then decompose the obtained expression in power of
$\frac{1}{n}$. Then
\[
F_{1}=e^{i(n+v)s}((p_{1}e^{is}-p_{-1}e^{-is})\frac{1}{\mid\delta\mid^{2}%
}(\frac{-1}{2n}+\frac{v}{2n^{2}}-\frac{4v^{2}+1}{8n^{3}}+O(\frac{1}{n^{4}}))+
\]%
\[
(p_{1}e^{is}+p_{-1}e^{-is})\frac{1}{\mid\delta\mid^{2}}(\frac{v}{4n^{2}}%
-\frac{v}{2n^{3}}+\frac{12v^{2}+1}{16n^{4}}+O(\frac{1}{n^{5}}))).
\]
Let $F_{2,1}$ and $F_{2,2}$ \ be the sum of terms in $F_{2}$ for which
$m+l=\pm2$ and $m+l=0$ respectively, i.e., $F_{2}=F_{2,1}+F_{2,2},$ where
\[
F_{2,1}=e^{i(n+v)s}(((p_{1})^{2}e^{2is}+(p_{-1})^{2}e^{-2is})\frac{1}%
{\mid\delta\mid^{4}}(\frac{-1}{8n^{2}}+\frac{-v}{4n^{3}}-\frac{12v^{2}%
+7}{32n^{4}}+O(\frac{1}{n^{5}}))+
\]%
\[
((p_{1})^{2}e^{2is}-(p_{-1})^{2}e^{-2is})\frac{1}{\mid\delta\mid^{4}}%
(\frac{-3}{16n^{3}}+O(\frac{1}{n^{4}}))),
\]%
\[
F_{2,2}=e^{i(n+v)s}\left\vert p_{1}\right\vert ^{2}(\frac{c_{20}}{n^{2}%
}++\frac{c_{21}}{n^{3}}+\frac{c_{22}}{n^{4}}+O(\frac{1}{n^{5}}))
\]
and $c_{20},c_{21},c_{22}$ are known constants. Similarly $F_{3}%
=F_{3,1}+F_{3,2}$, where $F_{3,1}$ and $F_{3,2}$ are the sum of terms in
$F_{3}$ for which $m+l+k=\pm3$ and $m+l+k=\pm1$ respectively. Hence
\[
F_{3,1}=e^{i(n+v)s}((p_{1}^{3}e^{3is}-p_{-1}^{3}e^{-is})\frac{1}{\mid
\delta\mid^{6}}(\frac{-1}{48n^{3}}+O(\frac{1}{n^{4}}))+
\]%
\[
(p_{1}^{3}e^{3is}+p_{-1}^{3}e^{-3is})\frac{1}{\mid\delta\mid^{6}}(\frac
{1}{16n^{4}}+O(\frac{1}{n^{5}}))),
\]%
\[
F_{3,2}=e^{i(n+v)s}((p_{1}e^{is}-p_{-1}e^{-is})\left\vert p_{1}\right\vert
^{2}(\frac{c_{23}}{n^{3}}+\frac{c_{24}}{n^{4}}+O(\frac{1}{n^{5}}))+
\]%
\[
(p_{1}e^{is}+p_{-1}e^{-is})\left\vert p_{1}\right\vert ^{2}(\frac{c_{25}%
}{n^{4}}+O(\frac{1}{n^{5}}))).
\]
In the same way we can write $F_{4}=F_{4,1}+F_{4,2}+F_{4,3}$, where $F_{4,1}$,
$F_{4,2}$, $F_{4,3}$ are the sum of terms in $F_{4}$ for which $m+l+k+r=\pm4$,
$m+l+k+r=\pm2$, $m+l+k+r=0$ respectively. Thus
\[
F_{4,1}=e^{i(n+v)s}(p_{1}^{4}e^{4is}+p_{-1}^{4}e^{-4is})\frac{1}{\mid
\delta\mid^{8}}(\frac{1}{384n^{4}}+O(\frac{1}{n^{5}})),
\]

\[
F_{4,2}=e^{i(n+v)s}(p_{1}^{2}e^{2is}+p_{-1}^{2}e^{-2is})\left\vert
p_{1}\right\vert ^{2}(\frac{c_{26}}{n^{4}}+O(\frac{1}{n^{5}}))),
\]

\[
F_{4,3}=e^{i(n+v)s}\left\vert p_{1}\right\vert ^{4}(\frac{c_{27}}{n^{4}%
}+O(\frac{1}{n^{5}}))).
\]
Since $p_{-1}^{k}e^{-iks}$ is conjugate of $p_{1}^{k}e^{iks}$ the real and
imaginary parts of $F_{k}e^{-i(n+v)s}$ consist of terms with multiplicands
$p_{1}^{k}e^{iks}+p_{-1}^{k}e^{-kis}$ and $p_{1}^{k}e^{iks}-p_{-1}^{k}%
e^{-iks}$ respectively. Taking into account this and using the above
estimations we get
\begin{align*}
|(\varphi_{n,v}(s),e^{i(n+v)s})\varphi_{n,v}(s)|^{2}  &  =2(\sum
_{k=1,2,3,4}\mathbf{Re(}F_{k})+\mathbf{Re(}F_{1}F_{2})+\mathbf{Re(}F_{1}%
F_{3}))\\
&  +|F_{1}|^{2}+|F_{2}|^{2}+O(n^{-5})=
\end{align*}%
\begin{align*}
&  1+\frac{1}{2n^{2}}\frac{1}{\mid\delta\mid^{2}}(p_{1}e^{is}+p_{-1}%
e^{-is}+c_{28}|p_{1}|^{2})+\frac{1}{n^{3}}((p_{1}e^{is}+p_{-1}e^{-is})c_{29}\\
&  +c_{30}|p_{1}|^{2})+\frac{1}{n^{4}}((p_{1}e^{is}+p_{-1}e^{-is}%
)c_{31}+c_{32}|p_{1}|^{2}+c_{33}|p_{1}|^{4}\\
&  +c_{34}|p_{1}|^{2}(p_{1}e^{is}+p_{-1}e^{-is})+(c_{35}+c_{36}|p_{1}%
|^{2})(p_{1}^{2}e^{2is}+p_{-1}^{2}e^{-2is}))+O(\frac{1}{n^{5}}),
\end{align*}
where $\mathbf{Re(}F)$ denotes the real part of $F.$ On the other hand
\[
|(\varphi_{n,v}(s),e^{i(n+v)s})|^{2}=(c_{37}\frac{1}{n^{2}}+c_{38}\frac
{1}{n^{3}}+c_{39}\frac{1}{n^{4}})|p_{1}|^{2}+c_{40}\frac{1}{n^{4}}|p_{1}%
|^{4}+O(\frac{1}{n^{5}}).
\]
The formula (19) follows from these equalities and (20) is a consequence of
(19), (17) and (15) for $k=2,4$ $\blacksquare$

\end{document}